\documentclass{IEEEtran}
\usepackage{amsmath}
\usepackage{listings}
\usepackage{times}
\usepackage[sort&compress,numbers]{natbib}
\usepackage{url}
\usepackage{graphicx}
\usepackage{graphics}
\usepackage{multirow}
\usepackage[normalsize]{subfigure}
\usepackage{wrapfig}
\usepackage[usenames,dvipsnames]{color}
\usepackage{multirow}
\usepackage{comment}

\newcommand{\Row}{R}
\newcommand{\Col}{C}
\newcommand{\scalar}{S}

\lstdefinestyle{basic}{showstringspaces=false,columns=flexible,language=Pascal,
escapechar=@,xleftmargin=1pc,%
mathescape=true,%
basicstyle=\sffamily,%
commentstyle=\mdseries,%
morekeywords={for,in,to,out,inout},%
deletekeywords={},%
moredelim=**[is][\color{white}]{~}{~},%
}
\begin{document}

\title{Reliable Generation of High-Performance\\ Matrix Algebra}

% I've put the authors in alphabetical order. -Jeremy

\author{
\IEEEauthorblockN{
Geoffrey Belter,\IEEEauthorrefmark{2}
Elizabeth Jessup,\IEEEauthorrefmark{3}
Thomas Nelson,\IEEEauthorrefmark{2}
Boyana Norris,\IEEEauthorrefmark{4} and
Jeremy G. Siek\IEEEauthorrefmark{2}}\\
\IEEEauthorblockA{\IEEEauthorrefmark{2}Department of Electrical, Computer, and Energy Engineering, University of Colorado, Boulder, CO 80309\\
{\small Email: [belter,Thomas.Nelson,Jeremy.Siek]@Colorado.EDU}}\\
\IEEEauthorblockA{\IEEEauthorrefmark{3}Department of Computer Science, University of Colorado, Boulder, CO 80309\\
{\small Email: jessup@Colorado.EDU}}\\
\IEEEauthorblockA{\IEEEauthorrefmark{4}Mathematics and Computer Science Division,
Argonne National Laboratory, Argonne, IL 60439\\
{\small Email: norris@mcs.anl.gov}}
}

\maketitle
%\IEEEpeerreviewmaketitle

% max 150 words
\begin{abstract}
Scientific programmers often turn to vendor-tuned Basic Linear Algebra
Subprograms (BLAS) to obtain portable high performance. However, many
numerical algorithms require several BLAS calls in sequence, and those
successive calls result in suboptimal performance.  The entire
sequence needs to be optimized in concert. Instead of vendor-tuned
BLAS, a programmer could start with source code in Fortran or C
(e.g., based on the Netlib BLAS) and use a state-of-the-art optimizing
compiler. However, our experiments show that optimizing compilers
often attain only one-quarter the performance of hand-optimized code.
In this paper we present a domain-specific compiler for matrix
algebra, the Build to Order BLAS (BTO), that reliably achieves
high performance using a scalable search algorithm for choosing the
best combination of loop fusion, array contraction, and multithreading
for data parallelism.  The BTO compiler generates code that is between
16\% slower and 39\% faster than hand-optimized code.

\end{abstract}

%\category{H.4}{Information Systems Applications}{Miscellaneous}  %TODO
%A category including the fourth, optional field follows...
%\category{D.2.8}{Software Engineering}{Metrics}[complexity measures, performance measures] %TODO

%\terms{Domain specific language, autotuning}

\section{Introduction}

Traditionally, scientific programmers have  used linear algebra libraries such
as the Basic Linear Algebra Subprograms (BLAS) \cite{Lawson,
  Dongarra88, Dongarra90} and the Linear Algebra PACKage (LAPACK)
\cite{Anderson} to perform their linear algebra
calculations.
A programmer links an application to vendor-tuned or autotuned
implementations of these libraries to achieve efficient,
portable applications.  For programs that rely on kernels with high
computational intensity, such as matrix-matrix multiply, this approach
can achieve near optimal performance~\cite{Whaley:1998fk}.  However,
memory bandwidth, not computational capacity, limits the performance
of many scientific applications~\cite{Anderson}, with data movement
expected to dominate the costs in the foreseeable
future~\cite{amarasinghe2009exascale}.

Because each BLAS performs a single mathematical operation,
such as matrix-vector multiplication, a tuned BLAS library has a
limited scope within which to optimize memory traffic.  
Moreover, separately compiled BLAS limit the scope of parallelization
on modern parallel architectures. Each BLAS call spawns threads and
must synchronize before returning, but much of this synchronization is
unnecessary when considering the entire sequence of matrix algebra
operations.
The BLAS Technical Forum suggested several new routines that combine
sequences of routines, thereby enabling a larger scope for
optimization~\cite{Blackford,Howell:2008:CEB:1356052.1356055}.
However, the number of useful BLAS combinations is larger than is
feasible to implement for each new architecture.
%% For
%% example, the four new kernels added into the BLAS standard are not
%% implemented in any of the vendor-tuned BLAS.
%% \geoff{MKL has introduced 2 api's they call 'blas like' one of which is in the new blas standard.  id also argue that without a netlib implementation its not safe for vendors to implement as there is no standard implementation for vendors to match.  the blast document does not describe behavior of corner cases.}

Instead of using vendor-optimized BLAS, a scientific programmer can start with
source code in Fortran or C, perhaps based on the Netlib BLAS,
and then use a state-of-the-art optimizing compiler to tune the code
for the architecture of interest. However, our experiments with two
industrial compilers (Intel and Portland Group) and one research
compiler (Pluto~\cite{Pluto}) show that, in many cases, these compilers achieve only
one-quarter of the performance of hand-optimized code (see
Section~\ref{sec:other_tools}). This result is surprising because the
benchmark programs we tested are sequences of nested loops with affine array
accesses, and the optimizations that we applied by hand 
(loop fusion, array contraction, and multithreading for data
parallelism) are well established.  Nevertheless, for some benchmarks, the
compilers fail to recognize
that an optimization is legal (semantics-preserving).  For other
benchmarks, they miscalculate the profitability of choosing one
combination of optimizations over another combination.

These observations demonstrate that achieving \emph{reliable},
automatic generation of high-performance matrix algebra is nontrivial. In
particular, the three main challenges are (1) recognizing whether an
optimization is legal, (2) accurately assessing the profitability of
optimizations and their parameters, and (3) efficiently searching
a large, discontinuous space of optimization choices and
parameters.  In this paper, we present the 
Build to Order BLAS (BTO) compiler.  It is the first compiler that solves
all three of these challenges specifically for the domain of \emph{matrix algebra}. 

%We emphasize that the scope of this work is \emph{matrix algebra}; not
%general-purpose compilation. 
%Many of the techniques that we use are
%specific to the domain of \emph{matrix algebra},
%but they may prove to be an important
%ingredient of a larger class of computations. Other
%domain-specific compilers, such as SPIRAL~\cite{Pueschel:05}, have
%already proven valuable in other domains (e.g., digital signal processing).
%

BTO accepts as input a sequence of matrix and vector
operations in a subset of MATLAB, together with a specification of the
storage formats for the inputs and outputs, and produces optimized
kernels in C. This choice of input language is part of our solution to
the problem of determining whether an optimization is legal. The
input language makes all data dependencies explicit, so there is no
difficulty recognizing whether an optimization is 
semantics-preserving or not.  Further, BTO uses a carefully designed internal
representation for optimization choices that rules out many
illegal choices while at the same time succinctly representing all the
legal choices. To accurately assess
profitability, the BTO compiler relies on a hybrid approach.
Analytic modeling is used for
coarse-grained pruning whereas empirical timing is used to make the
ultimate decisions. To find
the best combination of optimizations  within a large search space, BTO uses a genetic
algorithm whose initial population is the result of a greedy,
heuristic search.

We described earlier prototypes of BTO in several
papers~\cite{Siek,Belter,Belter2010,Karlin:2011:PMP:1964218.1964226}.
In these papers, we considered only a subset of the optimizations
considered here; moreover, at the time of their writing, we had not yet developed a search algorithm 
that was scalable with respect to the number of optimizations and
their parameters.
%
%% With respect to storage formats, BTO currently supports row-major and
%% column-major dense matrices. We plan to expand support to include
%% symmetric, triangular, banded, and sparse matrices.
%% %
%% Our first prototype of the BTO system fused loops at every
%% opportunity~\cite{Siek}.  The next refinement of the system added the
%% ability to explore all possible fusions and used a hybrid search
%% strategy that combined analytic modeling with empirical performance
%% testing~\cite{Belter}.  Later improvements to BTO included the
%% ability to produce shared memory parallel code \cite{Belter2010} and
%% analytically model shared memory parallel systems
%% \cite{Karlin:2011:PMP:1964218.1964226}.
%
%% While the hybrid exhaustive search is over a limited set of
%% parameters, it becomes impractical when searching over large numbers
%% of optimization parameters.  
%
%% In this paper, we present improvements to BTO's search capabilities
%% and present an analysis of various search strategies for fusion
%% combined with data parallelism.  In particular, we make the following
%% contributions:
%
%% In this paper we consider combinations of loop fusion, array
%% contraction, and multi-threading for data parallelism.  These three
%% optimizations are enough to outperform vendor-tuned BLAS and the
%% state-of-the-art optimizing compilers.
%
The following are the specific, technical contributions of this paper.

\begin{enumerate}
\item We present an internal representation for optimization choices
  that is \emph{complete} (includes all legal combinations of loop
  fusion, array contraction, and multithreading for data parallelism)
  but that inherently rules out many illegal combinations
  (Section~\ref{sec:search_description}).

\item We present a scalable and effective search strategy: a genetic
  algorithm with an initial population seeded by a greedy search.  We
  describe this strategy in Section~\ref{sec:search} and show 
  in Section~\ref{sec:other_tools} that it
  produces code that is between 16\% slower and 39\% faster than
  hand-optimized code.

\item We compare this genetic/greedy search strategy with several other
  strategies in order to reveal the rationale behind this strategy
  (Section~\ref{sec:compare-strategies}).
\end{enumerate}

%% In Section \ref{sec:bto}, we outline the workings of the BTO compiler
%% and describe the interface for plugging search strategies into the BTO
%% compiler.  In Section \ref{sec:results}, we discuss strategies for
%% searching through particular problems and define the search heuristics
%% for BTO.  We also present an empirical comparison of BTO to popular
%% linear algebra libraries for a test suite of kernels.  

We discuss related work in Section \ref{sec:related} and conclude the paper in
Section \ref{sec:conclusions} with a brief discussion of future work.

% LocalWords:  BLAS Dongarra PACKage LAPACK runtimes amarasinghe exascale BTO
% LocalWords:  Blackford unfused MATLAB Siek Belter BTO's Qasem Bondhugula
% LocalWords:  acache

\section{BTO Overview}
\label{sec:bto}
\label{sec:BTOFrameWork}

% The following is now redundant. We say this in the intro. -Jeremy
%% The BTO compiler takes a high-level description of a sequence of matrix algebra operations, in
%% a subset of MATLAB, and produces C code optimized for a particular
%% target architecture.  
Figure~\ref{fig:batax-spec} shows an example BTO input file for 
the BATAX subprogram that performs the operations $y \leftarrow \beta A^TAx$ for matrix $A$,
vectors $x$ and $y$, and scalar $\beta$.
The user of BTO specifies the input types, including storage formats
and a sequence of matrix, vector, and scalar operations; but the user
does not specify how the operations are to be implemented. That is,
the user does not identify such details as the
kinds of loops or the number of threads. The BTO compiler produces
a C implementation in two broad steps. It first chooses how to
implement the operations in terms of loops, maximizing
%being mindful to maximize
spatial locality by traversing memory via contiguous accesses.  It
then searches empirically for the combination of optimization decisions that
maximizes performance.  Sections \ref{sec:search_description} 
and \ref{sec:search_strat} describe the search process.

\begin{figure}[htb]
  \centering
%\lstset{language=MATLAB}
% with matlab set, beta gets bold face
\begin{small}
\begin{lstlisting}[label={lst:input},captionpos=b]
BATAX
in:
    x : vector(column), beta : scalar,
    A : matrix(row)
out: 
    y : vector(column)
{
    y = beta * A' * (A * x)
}
\end{lstlisting}
\end{small}
  \caption{BTO input file for the BATAX kernel.}
  \label{fig:batax-spec}
\end{figure}

\begin{figure}[htb]
\centering
\includegraphics[scale=0.5]{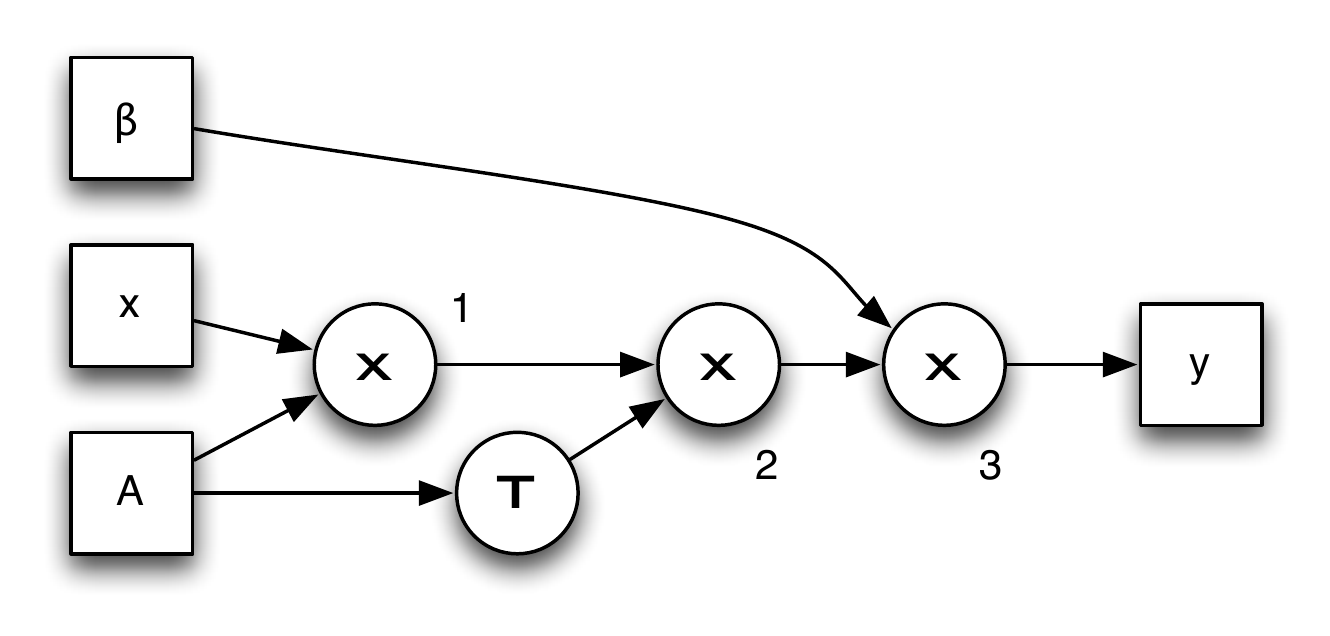}
\vspace{-15pt}
\caption{Dataflow graph for $y \leftarrow \beta  A^TAx$.}
\label{fig:atax_dataflow}
\end{figure}

Throughout the compilation process, BTO uses a dataflow graph
representation, illustrated in Figure \ref{fig:atax_dataflow} for the BATAX kernel. 
The square boxes correspond
to the input and output matrices and vectors, and the circles
correspond to the operations (operations are labeled with numbers, which are
used to identify the operations in the remainder of the paper).
%the numbers associated with each operation 
%are used later in the paper).

The BTO compiler uses a type system based on a container abstraction,
which describes the iteration space of matrices and vectors.  
Containers may be oriented horizontally or
vertically and can be nested.  We
assume that moving from one element to the next in a container is a
constant-time operation and good for spatial locality, but we place no
other restrictions on what memory layouts can be viewed as 
containers.  The types are defined by the following grammar, in which
$R$ designates row, $C$ designates column, and $S$ designates scalar.
\begin{equation*}
  \begin{array}{llcl}
    \text{orientations} & O & ::= & \Col \;|\; \Row \\
    \text{types}& T & ::= & O\texttt{<}T\texttt{>} \;|\; \scalar
  \end{array}
\end{equation*}

\begin{figure}[htb]
\centering
\includegraphics[scale=0.4]{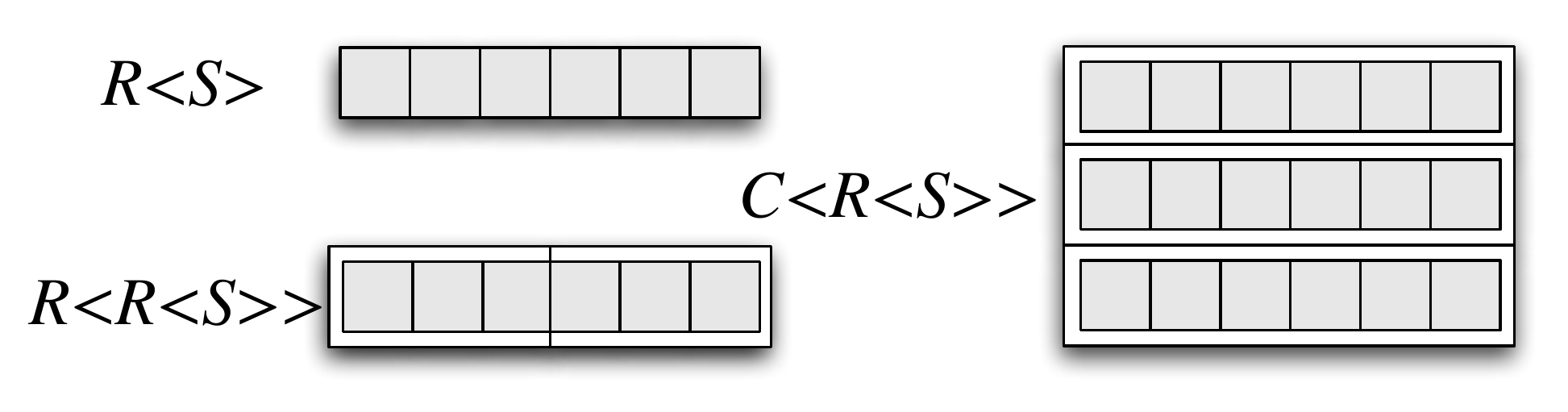}
\caption{Vector, partitioned vector, and matrix with their
  corresponding types.}
\label{fig:containers}
\end{figure}

Figure~\ref{fig:containers} shows several types with a corresponding
diagram depicting the container shapes: a row container with scalar
elements (upper left), a nested container for a row-major matrix
(right), and a partitioned row container (lower left).
%The nesting of containers
%enables the description of matrices, such as the row-major matrix
%on the right.  Additionally nesting can describe partitioned matrices and vectors.
%The lower-left diagram depicts a row that has been partitioned in half
%by adding an outer row container.  
%
During the creation of the dataflow graph, each node is assigned a
type.  The input and outputs are assigned types derived from the input
file specification, whereas the types associated with intermediate
results are inferred by the BTO compiler.

\paragraph*{Note on the polyhedral model} The type system used by BTO and
the polyhedral model~\cite{Karp:1967:OCU:321406.321418} share a common goal: 
both describe a schedule to
traverse an iteration space.  Much of BTO's functionality can be
accomplished by using polyhedral-based tools.  There are two motivations
for using a domain-specific type system as BTO does: (1) ability to
seamlessly perform additional optimizations (array contraction), and (2)
extensibility with regard to sparse matrix storage formats.

\section {Search Space}
\label{sec:search_description}

This section describes the search space and challenges with regard to
efficiently representing the space.  We present a domain specific
representation that enables BTO to eliminate many illegal points
without spending any search time on them.  This section also sets up
the discussion for specific search strategies in Section
\ref{sec:search_strat}.

\subsection{Description of Search Space}

The optimization search space we consider here has three dimensions: 
(1) loop fusion, (2) dimension or direction of data
partitioning, and (3) number of threads.  Even considering only these
three dimensions, there is a combinatorial explosion of optimization
combinations that BTO considers.  This search space is sparse, first
consisting of a high ratio of illegal compared to legal programs.
Within the legal programs, only a handful achieve good
performance.  The search space is also discrete because performance tends to
cluster with no continuity between clusters.  Efficiently searching
this space is the goal, and doing so requires a well-designed
representation.

Early versions of BTO's representation were too specific and therefore
limited the performance.  For example, they applied heuristics such as
fusing loops at every opportunity.  Experimental data show that in some
cases it is best to back off from full fusion, and the representation
needs to become more generic to accommodate that.
 
At the other extreme, we discovered that an overly generic
representation leads to the evaluation of an intractable number of
illegal versions.  For example, we tried a string-of-digits
representation that we describe in the next subsection. With this
approach, search time was dominated by the identification and
discarding of illegal programs.

\begin{figure}[htb]
\centering
\includegraphics[scale=0.5]{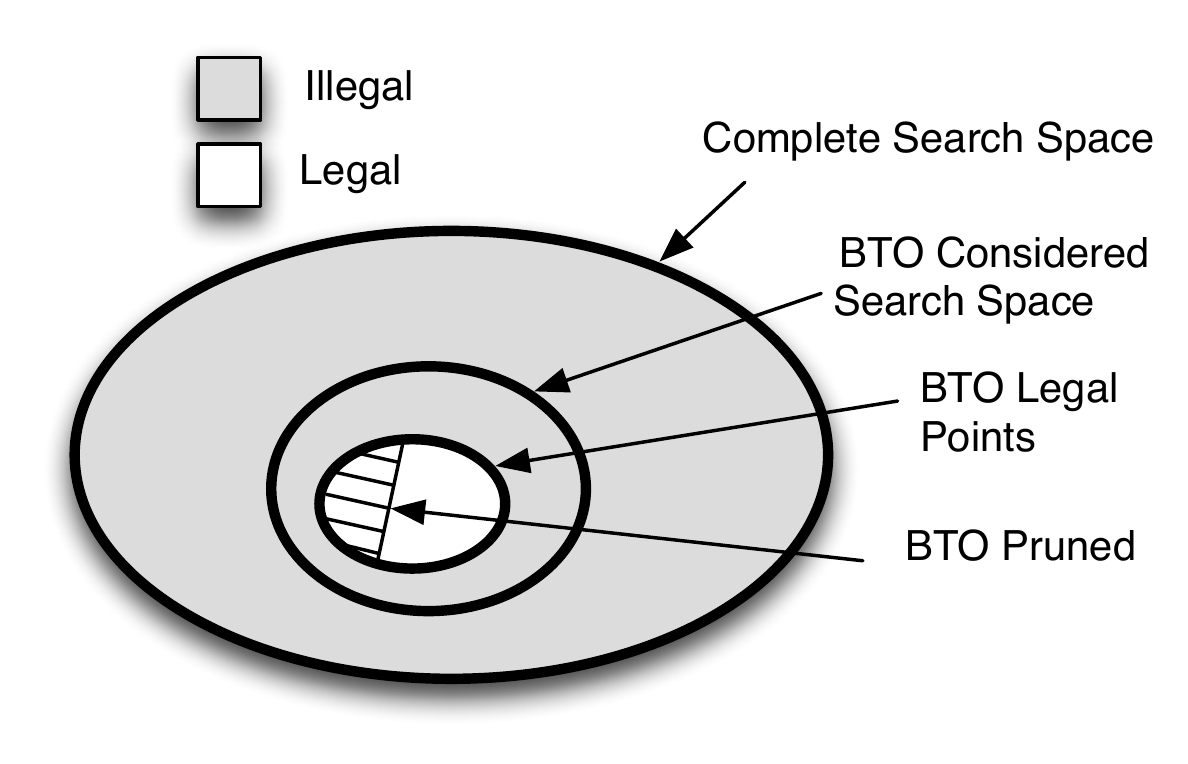}
\vspace{-15pt}
\caption{Representation of a typical search space showing how BTO avoids spending time searching a large portion of illegal points.}
\label{fig:searchspace}
\end{figure}

Figure \ref{fig:searchspace} shows a graphical representation of an
overly general search space and what area of that search space BTO
currently searches.  The gray areas represent illegal programs. This
area is large and, spending time  in it makes search times
intractable.  This sections describes a representation that allows
BTO to spend time only on the section labeled \emph{BTO Considered
  Search Space}, which contains many fewer illegal points.  Finally to
further improve search times, within the legal space, BTO prunes
points it deems unlikely to be unprofitable.  The rest of this
section walks through the findings that led to our current approach, as
well as the representation that enables a scalable search.

\subsection{Features of Search Space}

In an effort to interface with existing search tools, we initially
represented every fusion and parting decision in an easy to manipulate
set of digits.  For fusion we used an adjacency matrix that created
$((n-1)*n) / 2$ digits, where $n$ is the number of operations; 
partitions were represented with $2n$ digits, where each operation had
a direction of partition and a thread count.  As an example, a 
three-operation program would be represented as follows.
\begin{equation*}
f, f, f, w, t, w, t, w, t
\end{equation*}
\begin{equation*}
  \begin{array}{lcccr}
0 &\le &f &\le& max\_loop\_nest\_depth \\
1 &\le &w &\le& 3 \\
0 &\le &t &\le& max\_thread\_count 
  \end{array}
\end{equation*}
In the presence of one level of data partitioning and for a
matrix-vector type operation with a maximum thread count of 8, there are
over 1.2 million combinations of loop fusion and thread parallelism.

The primary advantage to this approach was that a search tool could
easily manipulate these strings of digits with no domain knowledge.
Unfortunately, search time was dominated by discarding illegal points.
Many of the illegal points were caused by a disrespect of interaction
between digits.  We now summarize two important features that this
representation does not encode.

\textbf{Fusion is an equivalence relation.} If an operation $a$ is fused with
operation $b$ and $b$ is fused with $c$, then $c$ must be fused with $a$.
Consider a three-operation program and the representation of fusion with an
adjacency matrix $M$, where $M[a,b]$ shows the depth of fusion between the loop
nests of $a$ and $b$.  Below, we show a valid fusion choice on the left and an
invalid fusion choice on the right. Each value in the matrix specifies fusing up
to two levels of nested loops. The matrix on the left describes fusing the outer
loop of all three operations, but only $b$ and $c$ have the inner loop fused.
The matrix on the right indicates fusing the inner loop of $a$ with $b$ and $b$
with $c$, but not $a$ and $c$, which of course is impossible.  We can describe
these constraints as forcing the relation specified in the adjacency matrix to be
an equivalence relation at every depth.
\begin{center}
\begin{minipage}{0.2\textwidth}
\[ \begin{array}{c|ccc}
& a & b & c  \\\hline
a & & 1 & 1 \\
b & &  & 2 \\
c & &  &  
\end{array} \] 
\end{minipage}
\begin{minipage}{0.2\textwidth}
\[ \begin{array}{c|ccc}
& a & b & c  \\\hline
a & & 2 & 1 \\
b & &  & 2 \\
c & &  &  
\end{array} \]
\end{minipage}
\end{center}

\textbf{Fused operations must use the same number of threads.}  Consider
a fuse graph that specifies a fusion of operations $a$ and $b$ but
then a partition that specifies $a$ use 4 threads and $b$ use 6
threads.  Partitioning the two operations with different thread counts
guarantees that fusion of these two operations is not possible.

Given the previous example program of three matrix-vector operation
with a maximum thread count of 8, respecting these features will bring the
possible points to consider down to a little over 1,000, or less than
one-tenth of a percent of points to consider without respecting these
features.

\subsection{Domain Specific Representation}
Designing a representation that respects the previously discussed
features requires domain knowledge.  At the expense of having to
custom-build search tools, we designed a representation to disallow,
with no search time required, a large number of illegal points.

Loop fusion is represented by \emph{fuse sets}.  Each operation is
given a unique identifier, and loops are represented with $\{\}$.  A
single loop operation (e.g., dot product) is represented as
$\{\mathit{ID}\}$, where $\mathit{ID}$ is a number identifying an
operation node in the dataflow graph. A two-loop operation, such as a
matrix-vector product, is represented as $\{\{\mathit{ID}\}\}$.  When
discussing a specific $\{\}$, we annotate it using $\{_i\}$, where $i$ describes
an axis of the iteration space.  We use $i$ to describe the iteration over
rows of a matrix and $j$ for columns of a matrix.  A complete
iteration space for a matrix can be described as $\{_i\{_j\}\}$ or
$\{_j\{_i\}\}$.

Fusion is
described by putting two operations within the $\{\}$. For example,
outer-loop fusion of two matrix-vector products is
described by $\{_j\{_i1\} \{_i2\}\}$, and fusion including the inner loops is
described by $\{_j\{_i1\; 2\}\}$.  
This notation encodes the equivalence
relation of loop fusion, disallowing a huge number of illegal fusion
combinations.

In BTO, fuse sets are actually more general than described in the
previous paragraph. In addition to representing loops, fuse sets can
represent iterations over tiles, spawning threads for data parallelism,
or loop unrolling. We refer to increasing the dimensionality of
the iteration space in this way as ``partitioning'' since it conceptually cuts a
matrix or vector into smaller parts.  A matrix-vector
operation of $\{_i\{_j1\}\}$ can be partitioned as $\{_{p(i)}\{_i\{_j1\}\}\}$
or $\{_{p(j)}\{_i\{_j1\}\}\}$, where the $\{\}$s 
annotated with $p(i)$ and $p(j)$ describe the new iteration dimension 
and the existing $i$ or $j$ loop variables that the partition affects.  The search tool
must specify which existing loop is being modified and how many
threads should be used.
The important
point here is that we can represent any level of nesting and describe
both C loops and data parallelism.
%By extending the fuse set representation to partitioning, there is a
%clean way to describe thread count in a way that respects the fusion
%choices.  
By extending the fuse set representation to partitioning,
thread counts can be assigned to each set, eliminating
the consideration of points with mismatched thread counts
within a fused operation.

BTO uses this representation to enumerate or manipulate the fuse sets 
and to generate the search space.  This
approach allows BTO to never touch the majority of the illegal
points it encountered with more general-purpose search tools.

\subsection {Discarding Remaining Illegal Points}
\label{sec:discard}
Recall Figure \ref{fig:searchspace} where the representation
applied by BTO reduces the search space to the area labeled \emph{BTO
  Considered Search Space}.  In this search space, a
significant number of illegal points remain.  Identifying them as
early as possible is key to a fast search.  This section describes how
BTO discards the remaining illegal points.
Figure~\ref{fig:atax_dataflow} shows the dataflow graph for the BATAX
operation $y
\leftarrow \beta A^TAx$
first described in Section~\ref{sec:bto}.  
Figure \ref{lst:atax_list} shows
each operation in BATAX 
numbered according to its corresponding
number in the dataflow graph.  Let us assume for
simplicity that subgraphs are fixed. Thus, although the scaling by
$\beta$ could be located differently in the graph, in this example it
cannot.

\begin{figure}[tb]
\begin{small}
\begin{lstlisting}[numbers=right,numberstyle=\small,numbersep=-50pt]
    t0 = A * x
    t1 = A' * t0
    y = t1 * beta
\end{lstlisting}
\end{small}
\vspace{-10pt}
\caption{Operation listing for $y \leftarrow \beta A^TAx$.}
\label{lst:atax_list}
\end{figure}

BTO performs type inference on the initial dataflow graph to check
whether the input program makes sense, assigning types to all
operations in the process. As BTO considers different optimization
choices, it incrementally updates the types to determine quickly
whether an optimization choice results in incompatible types.

In particular, illegal data dependency chains can be created with the
fuse set representation and therefore must be checked against the data
flow graph for correctness.  The following is a partial list of the
possible fuse sets for the running example.

\begin{small}
\begin{equation*}
\begin{array}{cc}
    a:  & \{\{1\}\} \{\{2\}\} \{\{3\}\} \\
    b:  & \{\{1\} \{2\}\} \{\{3\}\} \\
    c:  & \{\{1 2\}\} \{\{3\}\} \\
    d:  & \{\{1\} \{3\}\} \{\{2\}\} \\ 
    e:  & \{\{1\} \{2\} \{3\}\} \\
    f:   & \{\{1 2 3\}\} \\
\end{array}
\end{equation*}
\end{small}

Fuse set $d$ says to fuse operations 1 and 3. However, referring
to the dataflow graph in Figure~\ref{fig:atax_dataflow}, one can see
that there is a data dependency (operation 2) between 1 and 3.

A more subtle data dependency is caused by reduction operations.
Figure \ref{fig:atax_pseudo} shows the pseudocode for the example. 
Examination of the outer loops (lines 1 and 4) show that the
iterations are compatible and are legal to fuse.  Looking at the inner
loops (lines 2 and 5) we see compatible loops and assume fusion is
possible.  However, on line 3, \lstinline{t0[i]} is the destination of an
accumulation and is not available for use until the inner loop is
complete.  The next operation consumes this result and so the inner
loops cannot be fused.

 \begin{figure}[htb]
\begin{small}
\begin{lstlisting}[numbers=right,numberstyle=\tiny,numbersep=-30pt]
    for i in 1 to M
        for j in 1 to N
            t0[i] += A[i,j] * x[j]
    for i in 1 to M
        for j in 1 to N
            t1[j] += A[i,j] * t0[i]
    for j in 1 to N
        y[j] = t1[j] * beta
\end{lstlisting}
\end{small}
\caption{Pseudocode for unfused operations as shown in Figure~\ref{lst:atax_list}.}
\label{fig:atax_pseudo}
\end{figure}

The introduction of loops, the type inference, and the legality of
partition introduction are all based on the underlying type system
employed by BTO.  This system is described in detail in previous 
papers~\cite{Belter}.
Briefly, a set of rules describes legal linear algebra
operations based on the types involved in the operation.  Certain
rules cause a reduction, so an examination of the types involved in an
operation provides the loop nests and flags any loops as
performing a reduction.  In order to catch the reduction data dependency, data
flow analysis is combined with the result of examining the type to
determine that results are the destination of a reduction and that fusion cannot
occur.

The legality of every partitioning must also be checked for each
operation.
% i.e. $\{_{p(i)}\{_i\{_j1\}\}\}$ vs $\{_{p(j)}\{_i\{_j1\}\}\}$.  
In the absence of fusion, doing so is simply of a matter of
checking the type of each operand and the result of a given
operation.  The challenge is in identifying the set of partitions for each
operation such that fusion remains possible.  
The first operation of the BATAX example, $t0 = A \times x$, 
can be partitioned in the following ways.
\begin{equation*}
\begin{array}{lc|r}
(1) &t0(p) = A(p,:) \times x  \;\;\; &\;\;\; \{_{p(i)}\{_i\{_j1\}\}\}\\
(2) & t0 \,+\!=\, A(:,p) \times x(p) \;\;\; &\;\;\;  \{_{p(j)}\{_i\{_j1\}\}\}
\end{array}
\end{equation*}
Here we show the slicing of the matrix using the colon notation for a
complete iteration and $p$ for a subblock on which to operate in parallel
(borrowing notation from MATLAB). On the right is the representation
as a fuse set.  Partitioning (1) cuts the rows of $A$ and vector $t0$
while the second cuts the columns of $A$ and the vector $x$.
Partitioning (2) leads to a reduction at the parallel level, so $t0$
is not available for use until after a join.  The second operation of
the example, $t1 = A \times t0$ can be partitioned in the following ways.

\begin{equation*}
\begin{array}{lc|r}
(3) &t1(p) = A(p,:) \times t0 \;\;\; &\;\;\; \{_{p(j)}\{_i\{_j2\}\}\} \\
(4) & t1 \,+\!=\, A(:,p) \times t0(p) \;\;\; & \;\;\; \{_{p(i)}\{_i\{_j2\}\}\}
\end{array}
\end{equation*}

The question is how to partition operations 1 and 2 so that they 
can achieve fusion.  Data dependence analysis says that the
partition of operation 1, which introduces a reduction, will cause fusion
to fail, so operation 1 must be partitioned by using method (1) thus
limiting the options for operation 2.  The matrix $A$ is shared so, to
achieve fusion after partitioning, $A$ needs to be accessed the same way
in both partition loops.  From partitioning (1) we see that $A$ is accessed
as $A(p,:)$. Because operation 2 accesses the transpose of $A$, we must
select partitioning (4), accessing $A$ as $A(:,p)$.  By examining the
$\{\}$ notation we similarly observe that the partitions
introduced in (1) and (4) both generate $\{_{p(i)}\}$. In large fuse
sets, the likelihood of finding a correct set of operation
partitions randomly is small.

BTO uses a similar approach to that used in the BATAX example.  At the start of a
search, BTO enumerates the possible ways of partitioning for each
individual operation.  Then, when given a set of operations to fuse in
the presence of partitioning, a list of operation partitionings that
will allow fusion is found efficiently by comparing the shared data
structures in the operation (e.g., the matrix $A$ in BATAX).
This list may consist of zero to many combinations that work for a
fuse set, but all will be legal.  This approach quickly rules out the
illegal combinations, leaving only the legal points to consider.

\subsection{Discarding Unprofitable Points}
We again refer back to Figure \ref{fig:searchspace}, this time
considering the \emph{BTO Legal Points}, a small section
labeled \emph{BTO Pruned} represents legal points that typically
exhibit poor performance.  BTO uses a handful of heuristics to prune
these poorly performing points.  The first heuristic is to 
perform fusion only on operations that share an operand.  For example, if
one loop writes to a temporary matrix and another loop reads from the
temporary, then fusing the two loops reduces memory traffic.
Similarly, if two loops read from the same matrix, then fusion is
likely to be profitable. On the other hand, fusing loops that do not
share an operand is unlikely to reduce memory traffic.

%% Fusion of any loops will not be considered.  To
%% identify and discard disconnected loops requires analysis of the data
%% flow graph, however this can happen early in the process.

The next heuristic is that array contraction is always applied to
temporary data structures in the presence of fusion. Again, reducing
memory traffic almost always improves performance.
%% Fusion of loops
%% without eliminating the extra memory traffic may have less of an
%% effect on performance.

%% The third heuristic is to ensure that inner-most loops always traverse
%% matrices along contiguous memory.  For example the inner loop over a
%% column-major matrix always traverses over a column.

The second two heuristics eliminate points without having to spend any
time on those that are unprofitable.  The array contraction is always
performed while the contiguous traversal is encoded in the type system
exploited by BTO.

%% The second two heuristics eliminate points without having to spend any
%% time on those that are unprofitable.  The array contraction is always
%% performed while the contiguous traversal is encoded in the type system
%% utilized by BTO.

\section{Genetic/Greedy Search Strategy}
\label{sec:search}
\label{sec:search_strat}

This section describes the BTO search strategy based on a genetic
algorithm whose initial population is determined by a greedy search
that tries to maximally fuse loops. We refer to this search strategy
as MFGA, for Maximal Fusion followed by Genetic Algorithm. Section
\ref{sec:results} explores why this search is used, and the value of
heuristics and alternatives.

We explain MFGA using the \( y \leftarrow \beta A^TAx\)  BATAX example
from the previous section.  Genetic algorithms are a broad category of global
optimization techniques inspired by biological evolution
\cite{mitchell1998introduction}. In genetic algorithms,
each code version is called an organism.  A genetic algorithm uses a population
of organisms. At each generation, the worst organisms are removed from the
population and are replaced with newly generated organisms.
%
%A high-level outline of the GA algorithm is given in Figure~\ref{code:GA-Algorithm}.
%\boyana{I think the GA pseudocode is too high-level and generic to be of great value, people who know anything about genetic algs won't get much out of it, and there isn't enough to educate those who don't. I suggest removing it.}
%\begin{figure}
%\begin{scriptsize}
%\begin{lstlisting}
%population = N random organisms (Max-Fuse+Mutation)
%while (!finished):
%	test population; fitness=performance 
%	choose 2*N parents using fitness
%	population = N new organisms with crossover 
%	randomly mutate some children
%\end{lstlisting}
%\end{scriptsize}
%\caption{Pseudo-Code for the GA algorithm}
%\label{code:GA-Algorithm}
%\end{figure}

\subsection{Max Fuse}

The search begins with a greedy Max-Fuse (MF) heuristic: we attempt to
fuse as many of the loops as possible to the greatest depth possible,
subject to the constraints described in Section
\ref{sec:search_description}.
%% As described in that section, The
%% compiler views loop fusion and fusion of thread operations similarly,
%% based on operation types (\thomas{ Make sure we actually describe
%%   this!}).  Thus the two matrix-vector products in our example have a
%% total possible ``fusion depth'' of three: the two c loops, and the
%% outer thread dispatch.  \boyana{What are the "c" loops?}  We can
%% represent this as the following nested sets of operations:
The MF search starts from unfused but partitioned versions of the
kernel in which the axis of partitioning has not yet been
decided. Continuing with the BATAX example from the previous
section, the following represents the
unfused partitioned kernel.  The $X$, $Y$, and $Z$ are unknowns
determined during the MF search.
\begin{gather*}
\{_X\{_i\{_j 1\}\}\} \{_Y\{_i\{_j 2\}\}\} \{_Z\{_j 3\}\} 
\end{gather*}
To fuse the $X$ and $Y$ iterations, we need $X = Y$, so we proceed
with the fusion and constrain ourselves to $X = Y$.
\begin{align*}
 & \{_X\{_i\{_j 1\}\}\} \{_X\{_i\{_j 2\}\}\} \{_Z\{_j 3\}\} \\
\Rightarrow\; & \{_X\{_i\{_j 1\}\} \{_i\{_j 2\}\}\} \{_Z\{_j 3\}\} 
\end{align*}
At this point, $X$ has to be $p(i)$ because the alternative,
$p(j)$, would mean that the necessary results from operation 1 would
not be available for operation 2.
Next, we can also fuse the $i$ iteration of operations 1 and 2.
\begin{align*}
 & \{_{p(i)}\{_i\{_j 1\}\} \{_i\{_j 2\}\}\} \{_Z\{_j 3\}\} \\
\Rightarrow\; & \{_{p(i)}\{_i\{_j 1\} \{_j 2\}\}\} \{_Z\{_j 3\}\} 
\end{align*}
Because of the reduction in the matrix-vector product (operation 1),
the $j$ iteration of operations 1 and 2 cannot be fused.

Next, we consider whether the $p(i)$ iteration can be fused with $Z$.  The
$p(i)$ iteration requires a reduction before the final vector scaling
of operation 3, so 3 must reside in its own thread.  Finally, there is
only one axis of iteration in operation 3, so $Z$ must be $p(j)$.
Therefore, the MF search produces the following organism: $\{_{p(i)}
\{_i \{_j 1\} \{_j 2\} \}\} \{_{p(j)}\{_j 3\}\}$.

%% Recall from the previous section there is only one possible data
%% partitioning that allows both operations to occur in the same thread:
%% part of Max-Fuse is pruning the space to find the legal partition
%% \emph{ways}. \thomas{Should this be explained here or in Search
%%   Space?}

\subsection{Generation of Initial Population by Mutation}

The initial population for the genetic algorithm is created by
applying random mutation to the Max-Fuse point.
%
%% These mutations can also be applied with some small probability after
%% each crossover to increase population diversity.
%
Each mutation performs \emph{one} of the following four changes:
(1) add or remove fusion level,
(2) add or remove partition level,
(3) change the partition axis, or
(4) change the number of threads.
Mutations are constrained to the set of legal organisms; for example,
attempting to add a mutation to the already maximally fused point from our
previous example will fail, resulting in no change. However, mutations might randomly
remove the partition from operation 3: 
\[
\{_{p(i)}\{_i \{_j 1\} \{_j 2\} \}\} \{_{p(j)}\{_j 3\}\}
\Rightarrow
\{_{p(i)}\{_i \{_j 1\} \{_j 2\} \}\} \{_j 3\}
\]

The Random search described in Section~\ref{sec:fuseperf} consists of
repeatedly applying random mutations to the organism without any
further search.

\subsection{Selection and Crossover}

After the initial generation of organisms (and for every following generation), 
we compile and test every organism and record its runtime,
which is the value the search tries to minimize.  We then select \(2
N\) of the fittest organisms to be parents for the next generation,
where the population size $N$ can be user specified, but defaults to
$20$.

\paragraph{Parent Selection Method}

The population evolves via tournament
selection~\cite{mitchell1998introduction}: $k$ random organisms are
chosen to be potential parents, and the potential parent with the best
fitness becomes an actual parent.  This process balances hill climbing with
exploration, allowing less fit organisms to sometimes become parents,
and thus helping the algorithm escape locally optimal solutions that are
not globally optimal.  Larger values of $k$ cause the algorithm to
converge more quickly on a solution, while smaller values of $k$
converge more slowly but increases exploration. BTO uses $k=2$ to
favor exploration.

\paragraph{Crossover}

Crossover is a function that takes two parent organisms and randomly
chooses features of the two parents to create a child organism.  The
key strength of genetic algorithms is that crossover can sometimes
combine the strengths of two versions.  Our crossover function
generates the child recursively from the two parents, making fusion
decisions at each level and making sure those decisions remain valid
for inner levels.

Our crossover function uses the type-based representation of each
expression as described in Section~\ref{sec:bto} and performs
crossover by comparing the two types.
Continuing with the \( y \leftarrow \beta A^TAx\) example, consider
the following two organisms $A$ and $B$.
\begin{align*}
A:\;& \{_{p(i)}\{_i \{_j 1\} \{_j 2\} \}\} \{_j 3\} \\
B:\;& \{_{p(i)}\{_i\{_j 1\}\}\} \{_{p(i)}\{_i\{_j 2\}\}\} \{_{p(j)}\{_j 3\}\} 
\end{align*}
Parent A partitions and partly fuses operations 1 and 2 but does not
partition operation 3.  Parent B has all partitions turned on but has
not fused operations 1 and 2.

Crossover chooses which parent to emulate for each operation, working
from the outermost fuse level inwards.  Each step constrains the
possibilities for the other operations.  In our example, crossover
might choose parent A for the outermost level of operation 1, meaning
1 and 2 exist in the same thread (also using Parent A's
partitioning axis $p(i)$ and thread number data).  It then might
choose Parent B for the next level, iteration $i$.  This mechanism
forces operation 1 and 2 not to be fused, resulting in
$\{_i\{_j 1\}\} \{_i\{_j 2\}\}$.
Then the crossover moves to operation 3 and the process continues.  If
B is chosen, the final child becomes $\{_{p(i)} \{_i \{_j 1\}\} \{_i
\{_j 2\}\} \} \{_{p(j)} \{_j 3\}\}$.

The tournament selection process is repeated \(N\) times, creating a
new generation of organisms.  Fitness values are cached. If crossover
ever produces an organism that was already tested in a previous
generation, the old value is used to save search time.

\subsection{Search for Number of Threads}
\label{sec:thread-perf}
% initial: # cores
% 

BTO uses a fixed number of threads to execute all of the data-parallel
partitions in a kernel. We refer to this as the \emph{global thread
  number} heuristic.  An alternative is to allow different
numbers of threads for each partition, which we refer to as the
\emph{exhaustive thread} search.  In Section~\ref{sec:threadperf}, we
present data that show that the exhaustive approach takes much more
time but does not lead to significantly better results.

BTO includes the search for the best number of threads in the MFGA
algorithm. The initial number is set to the number of cores in the
target computer architecture. The mutation function either increments
or decrements the thread number of 2. The crossover function simply
picks the thread number from one of the parents.  After the genetic
algorithm completes, MFGA performs an additional search for the best
number of threads by testing the performance when using thread counts 
between 2 and the number of cores, incrementing by 2.

%% Our thread search uses what we call the \emph{global thread number}
%% heuristic.  The global thread number specifies how many threads every
%% data-parallel operation should spawn, i.e., into how many pieces to
%% partition the data.  Using our example organism $\{\{ \{1\} \{2\} \}\}
%% \{\{3\}\}$, operations 1 and 2 are in a thread together, and operation
%% 3 is in its own thread.  BTO can use different partition sizes for
%% each of those thread operations.  The heuristic is to test only
%% versions where the same number of threads is spawned for the 1-2
%% operation as for the 3 operation. Our hypothesis that this improves
%% scalability without sacrificing performance is tested in
%% \ref{sec:threadperf}.

\section{Results}
\label{sec:results}

We begin this section with a comparison of the performance of
BTO-generated routines and several state-of-the-art tools and
libraries that perform similar sets of optimizations, as well as
hand-optimized code.  The BTO compiler generates code that is between
16\% slower and 39\% faster than hand-optimized code.
The other automated tools and libraries achieve comparable performance
to BTO and hand-optimized code for only a few of the kernels.
For the smaller kernels in which we can exhaustively enumerate all
possible combinations of optimizations, we show that the MFGA search
method finds a routine that performs within 2\%
of the best found in the exhaustive search.
%
%Finally 
We present empirical data that motivates our choice of the
MFGA search strategy, comparing MFGA to several alternative strategies
and analyzing the orthogonality of fusion choices versus number of
threads.

\subsection {Test Environment and Kernels}

The results in this section rely on the kernels shown in Table
\ref{tab:routines}.  Some of these kernels respond well to loop fusion
and data parallelism while others do not.  Some of these kernels are
in the updated BLAS~\cite{Blackford} but have not been adopted by
vendor-tuned BLAS libraries.  These kernels also represent various
uses of the BLAS.  For example, the DGEMV kernel maps directly to a
BLAS call while others are equivalent to multiple BLAS calls.  As an
example, Listing \ref{fig:bicg-blas} shows the sequence of BLAS calls
that implement the BICGK kernel.

\begin{table}[h]
\caption{Kernel specifications.}
\vspace{-10pt}
\label{tab:routines}
\begin{center}
\begin{tabular}{|c|c|}
\hline
\textbf{Kernel} & \textbf{Operation} \\ \hline
\multirow{2}{*}{AXPYDOT} & $ z \gets w - \alpha v$ \\
  &       $\beta \gets z^Tu$  \\  \hline
VADD & $x \gets w + y + z$ \\ \hline
WAXPBY  & $w \gets \alpha x + \beta y$\\ \hline\hline
ATAX   & $y \gets A^TAx$ \\  \hline
\multirow{2}{*}{BICGK} & $q \gets Ap$ \\
  &          $s \gets A^Tr$ \\ \hline
DGEMV  & $z \gets \alpha Ax + \beta y$ \\ \hline
\multirow{2}{*}{DGEMVT}  & $x \gets \beta A^Ty + z$ \\
  &       $w \gets \alpha Ax$ \\ \hline
\multirow{3}{*}{GEMVER} & $B \gets A + u_1v_1^T + u_2v_2^T$ \\
& $x \gets \beta B^Ty + z$ \\
& $w \gets \alpha Bx$ \\ \hline
GESUMMV & $y \gets \alpha Ax + \beta Bx$ \\ \hline
\end{tabular}
\end{center}
%\vspace{-10pt}
\end{table}

\begin{figure}[tb]
  \centering
\begin{small}
\begin{lstlisting}[label={lst:bicgk},captionpos=b]
// q = A * p
dgemv('N',A_nrows,A_ncols,1.0,A,lda,p,1,0.0,q,1);
// s = A' * r
dgemv('T',A_nrows,A_ncols,1.0,A,lda,r,1,0.0,s,1);
\end{lstlisting}
\end{small}
  \caption{Example sequence of BLAS calls that implement BICGK.}
  \label{fig:bicg-blas}
\end{figure}
                
The computers used for testing include recent AMD and Intel multicore
architectures which we describe in Table \ref{tab:specs}.

\begin{table}[h]
\caption{Specifications of the test machines.}
\vspace{-10pt}
\label{tab:specs}
\begin{center}
\begin{tabular}{|@{\hspace{1pt}}p{58pt}@{\hspace{1pt}}|c|c|c|c|c|}
\hline
\textbf{Processor} & \textbf{Cores} & \textbf{Speed} & \textbf{L1} & \textbf{L2} & \textbf{L3}\\ 
 &  & \textbf{(GHz)} & \textbf{(KB)} & \textbf{(KB)} & \textbf{(MB)}\\ 
\hline
{\scriptsize Intel Westmere}& 24 & 2.66 & 12 x 32 & 12 x 256 & 2 x 12 \\ \hline
%Westmere  & & & & &  \\ \hline
{\scriptsize AMD Phenom II X6} &  6 & 3.3 & 6 x 64 &  6 x 512 & 1 x 6 \\ \hline
% Phenom II & & & & & \\ 
%  X6 & & & & & \\ \hline
{\scriptsize AMD Interlagos} & 64 & 2.2 & 64 x 16 & 16 x 2048 & 8 x 8 \\
%  Interlagos & & & & & \\
\hline
\end{tabular}
\end{center}
\end{table}

\subsection {Comparison to Similar Tools}
\label{sec:other_tools}

We begin by placing BTO performance results in context by comparing them 
with several state-of-the-art tools and libraries.  BTO performs loop
fusion and array contraction and makes use of data parallelism.  BTO
relies on the native compiler for loop unrolling and vectorization.
The two compilers used to gather this data are the Intel C Compiler
(ICC)~\cite{icc} and the PGI C Compiler (PGCC)~\cite{pgi}.  With the
exception of the explicitly labeled PGCC results, all kernels are
compiled using ICC.  Both ICC and PGCC unroll loops and vectorize.
They also identify and exploit data parallelism and perform loop fusion.
%\liz{took out claim that they do this stuff well since we complain
%about them soon after}

We begin with a detailed comparison of BTO and
five other approaches for generating high performance code on the
Intel Westmere. We then give a brief summary of similar results on the
AMD Phenom and Interlagos.

The first approaches are using ICC and PGCC, which represent the best
commercial compilers.
The third approach is using Pluto~\cite{Pluto}, a source-to-source
translator capable of performing loop fusion and identifying data
parallelism.  The fourth approach is using Intel's Math Kernel Library
(MKL)~\cite{icc} which is a vendor-tuned BLAS implementation targeting
Intel CPUs.  The fifth is a hand-tuned implementation (applying loop
fusion, array contraction, and data parallelism) created by 
%the first author, who is 
an expert in performance tuning who works in the
performance library group at Apple, Inc.  The input for ICC, PGCC and
Pluto is a translation of the BLAS calls to C loops.
%, so the only
%fusion in the input program is limited to that found in BLAS.  
The compiler flags 
used with ICC were \emph{``-O3 -mkl -fno-alias"} and the flags for PGCC 
were \emph{``-O4 -fast -Mipa=fast -Mconcur -Mvect=fuse -Msafeptr"} 
(\emph{``-Msafeptr"} not used on Interlagos). Data parallelism is 
exploited by ICC, PGCC, Pluto,
and MKL by using OpenMP \cite{Dagum:1998:OIA:615255.615542}.  BTO and the hand-tuned
versions use Pthreads \cite{Mueller94pthreadslibrary}.

%% BN: moving table contents to text (for space reduction).
%flags used for compilation are shown in Table \ref{tab:flags}.
%
%\begin{table}[tb]
%\caption{Compiler flags. Msafeptr not used on Interlagos).}
%\vspace{-10pt}
%\label{tab:flags}
%\begin{center}
%\begin{tabular}{|c|c|}
%\hline
%ICC & \emph{-O3 -mkl -fno-alias} \\ \hline
%PGCC & \emph{-O4 -fast -Mipa=fast -Mconcur -Mvect=fuse -Msafeptr} \\ \hline
%\end{tabular}
%\end{center}
%\end{table}%

\begin{figure*}[tbp] 
  \includegraphics[width=\textwidth]{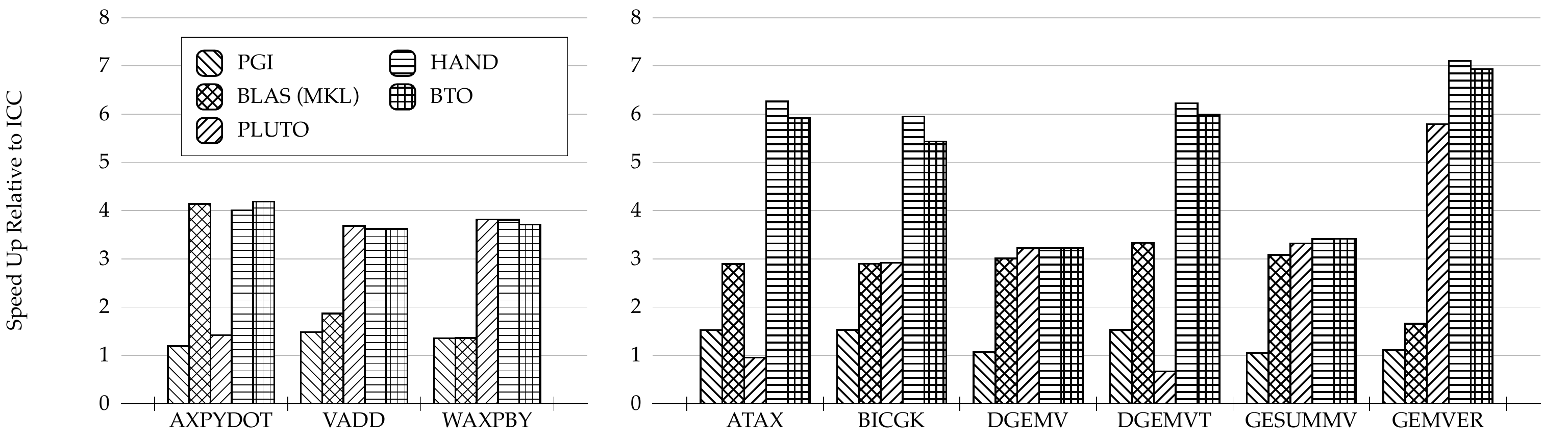} 
   \caption{Performance data for Intel Westmere.  Speedups relative to unfused loops compiled with ICC (ICC performance is 1 and not shown).  The left three kernels are vector-vector while the right six are matrix-vector operations. In all cases, BTO generates code that is between
16\% slower and 39\% faster than hand-optimized code and significantly faster than library and compiler-optimized versions.}
   \label{fig:bto_compare}
\end{figure*}

Figure \ref{fig:bto_compare} shows the speedup relative to ICC on the
y-axis for several linear algebra kernels.  (ICC performance
is 1.) On the left are the three vector-vector kernels, and on the
right are the six matrix-vector kernels, all from Table
\ref{tab:routines}.

PGCC tends to do slightly better than ICC, with speedups ranging from
1.1 to 1.5 times faster.  Examining the output of PGCC shows that all
but GESUMMV and GEMVER were parallelized.  However, PGCC's ability to
perform loop fusion was mixed; it fused the appropriate loops in
AXPYDOT, VADD, and WAXPBY but complained of a ``complex flow graph''
on the remaining kernels and only achieved limited fusion.

The MKL BLAS outperforms ICC by factors ranging from 1.4 to 4.2.  The
calls to BLAS routines prevent loop fusion, so significant
speedups, such as those observed in AXPYDOT and GESUMMV, can instead
be attributed to parallelism and well tuned vector implementations of
the individual operations.
%Points of note are AXPYDOT and GESUMMV which do
%quite well.  
%GESUMMV has limited benefit from loop fusion, implying the optimizations
%performed within the BLAS are enough to be competitive. 
We were unable to determine why the BLAS perform so well for AXPYDOT.
Surprisingly, the BLAS DGEMV does not perform as well as Pluto and BTO.
% which is surprising, considering  this is an
%exact BLAS call.  
Given the lack of fusion potential in this kernel, we
speculate that differences in the parallel implementations are the cause.  

The Pluto results show speedups ranging from 0.7 to 5.7 times faster
than ICC.  The worst-performing kernels are AXPYDOT, ATAX, and DGEMVT.
These three kernels represent the only cases where Pluto did not
introduce data parallelism.  For the remaining two vector-vector
kernels, VADD and WAXPBY, Pluto created the best-performing result,
slightly better than the BTO and hand-tuned versions.  Inspection
shows that the only difference between Pluto, hand-tuned, and BTO in
these cases was the use of OpenMP for Pluto and Pthreads for
hand-tuned and BTO.  The fusion was otherwise identical and the
difference in thread count had little effect.  For the matrix-vector
operations, if we enabled fusion but not parallelization with Pluto's
flags, then Pluto matched BTO with respect to fusion. However, with
both fusion and parallelization enabled, Pluto sometimes misses
fusion and/or parallelization opportunities.  For example BICGK was
parallelized but not fused.  The GEMVER results depend on the loop
ordering in the input file. For GEMVER, Pluto performed either
complete fusion with no parallelism or incomplete fusion with
parallelism; the latter provided the best performance and is shown in
Figure~\ref{fig:bto_compare}.

The hand-tuned implementation is intended as a sanity check.  For the
vector-vector operations, the hand-tuned version is within a few
percent of the best implementation. Typically the fusion in both the hand tuned
and the best tool based version are 
identical with the primary difference being either thread count or
what appears to be a difference between Pthreads and OpenMP
performance.  In the case of the matrix-vector operations, the
hand-tuned version is the best for all but DGEMV and GESUMMV, where it
is equal to the best.

The BTO performance results show speedups ranging from 3.2 to 6.9
times faster than ICC.  For the vector-vector operations, the
performance is similar to the hand-tuned version in all cases.
Inspection shows that for AXPYDOT, BTO was slightly faster than the
hand-tuned version because BTO did not fuse the inner loop while the
hand-tuned version did. BTO performed slightly worse than hand-tuned
on WAXPBY because of a difference in thread count.  Similarly, the
performance of the matrix-vector operations is close but slightly
lower than that of the hand-tuned version.  BTO fused the same as
hand-tuned for BICGK, GEMVER and DGEMVT with the only difference being
in thread count.  For ATAX, both BTO and hand-tuned fused the same and
selected the same number of threads, but BTO was slightly slower
because of where it zeroed out a data structure. In the hand-tuned
version the zeroing occurred in the threads, while in BTO's case it
occurred in the main thread.

%THIS IS FIG 7

Similar results on AMD Phenom and AMD Interlagos are shown in Table
\ref{tbl:bto_compare_amd} and
Table~\ref{tbl:bto_compare_amd_interlagos}, respectively.
%The results are similar to those discussed with the exception of the Pluto results.  
The Pluto-generated code for the matrix-vector operations tended to perform worse than that produced for 
the other methods evaluated.  On this computer, achieving full fusion while maintaining 
parallelism is of great importance. As previously discussed, Pluto tended to achieve fusion or 
parallelism but struggled with the combination. These results demonstrate the difficulty of portable 
high-performance code generation even under autotuning scenarios.  

\begin{table}[tb]
\caption{Performance data for AMD Phenom.  BLAS numbers from AMD's ACML.
Speedups relative to unfused loops compiled with PGCC (PGCC performance is 1 and
not shown). Best performing version in bold.}
\begin{center}
\begin{tabular}{|c|c|c|c|c|}
\hline
\textbf{Kernel} & \textbf{BLAS} & \textbf{Pluto} & \textbf{HAND} & \textbf{BTO} \\ \hline
AXPYDOT & 0.97 & 1.81 & 1.58 & \textbf{1.86} \\  \hline
VADD & 0.84 & 1.33 & 1.50 & \textbf{1.83} \\ \hline
WAXPBY  & 0.79 & 1.40 & 1.68 & \textbf{1.91} \\ \hline \hline
ATAX   & 1.27 & 0.69 & \textbf{2.92} & \textbf{2.92} \\  \hline
BICGK & 1.27 & 0.80 & 2.80 & \textbf{2.84}\\ \hline
DGEMV  & 1.67 & 0.71 & 1.85 & \textbf{1.89} \\ \hline
DGEMVT  & 1.67 & 0.71 & 1.85 & \textbf{1.89} \\ \hline
GEMVER & 1.04 & 1.61 & \textbf{2.61} & 2.34 \\ \hline
GESUMMV & 1.63 & 0.63 & 1.74 & \textbf{1.75} \\ \hline

\hline
\end{tabular}
\end{center}
\label{tbl:bto_compare_amd}
\end{table}

\begin{table}[tb]
\caption{Performance data for AMD Interlagos.  BLAS numbers from AMD's ACML.  Speedups relative to unfused loops compiled with PGCC (PGCC performance is 1 and not shown). Best performing version in bold.}
\begin{center}
\begin{tabular}{|c|c|c|c|c|}
\hline
\textbf{Kernel} & \textbf{BLAS} & \textbf{Pluto} & \textbf{HAND} & \textbf{BTO} \\ \hline
AXPYDOT &  0.82 & 1.60 & 1.73 & \textbf{1.61} \\  \hline
VADD &  0.43 & 1.05 & 1.14 & \textbf{1.15} \\ \hline
WAXPBY  & 0.34 & 1.06 & \textbf{1.16} & 1.11 \\ \hline \hline
ATAX   & 2.49 & 0.43 & 4.09 & \textbf{4.28} \\  \hline
BICGK & 2.35 & 1.60 & 3.03 & \textbf{4.22} \\ \hline
DGEMV  & \textbf{2.45} & 0.89 & 1.66 & 2.07 \\ \hline
DGEMVT  & 2.43 & 0.43 & \textbf{4.08} & 4.03 \\ \hline
GEMVER & 1.70 & 2.00 & \textbf{4.15} & 4.05 \\ \hline
GESUMMV & \textbf{2.36} & 0.37 & 1.65 & 2.03 \\ \hline

\hline
\end{tabular}
\end{center}
\label{tbl:bto_compare_amd_interlagos}
\end{table}

%%%%%% if we want to use a graph instead of a table its here %%%%%%%%
%\begin{figure*}[tb] 
%  \includegraphics[width=\textwidth]{amd_compare.pdf} 
 %  \caption{Performance data for AMD Phenom.  Speedups relative to unfused loops compiled with PGCC (PGCC performance is 1 and not shown).  The left three kernels are vector-vector while the right six are matrix-vector operations.}
%   \label{fig:bto_compare_amd}
%\end{figure*}

\paragraph*{Summary}

Compared with the best alternative approach for a given kernel, BTO
performance ranges from 16\% slower to 39\% faster.  Excluding
hand-written comparison points, BTO performs between 14\% worse and
229\% better.  Pluto, ICC, PGCC, and BLAS all achieve near-best
performance for only a few points; however, BTO's performance is most
consistent across kernels and computers. Excluding the hand-optimized
results, BTO finds the best version for 7 of 9 kernels on the Intel
Westmere, all 9 kernels on the AMD Phenom, and 7 of 9 kernels on the AMD
Interlagos. Surprisingly, on the AMD Phenom, BTO surpassed the
hand-optimized code for 7 of the 9 kernels and tied for one kernel.

\subsection{MFGA Compared to Exhaustive Searches}

In Section \ref{sec:other_tools}, we presented results showing that
BTO's MFGA search strategy finds high-performing versions for a range
of kernels. In this section, we show how the performance of the MFGA
search strategy compares with the best version that can be produced
using exhaustive or nearly exhaustive search strategies on Intel
Westmere.  These strategies require long-running searches that can
take days to complete. For the smaller kernels, a completely
exhaustive search is possible. For larger kernels, exhaustive search
was not feasible, so we instead use a strategy that is exhaustive with
respect to each optimization, but orthogonal between
optimizations. For the largest kernels, GEMVER and DGEMV, even the
orthogonal approach took too much time, not completing even after
weeks of running.
%% Kernels fall into
%% three categories when it comes to long running searches: exhaustive
%% coverage is possible, broad coverage is possible, or the
%% %search fails due to the size of the search space.
%% coverage remains too sparse after weeks of running.
%% For the first two categories, there is a point of comparison for the fast-running search
%% strategy. 

%Table \ref{tab:percentile} shows 
We compared the performance of kernels produced by MFGA as percentage
of the exhaustive search for smaller kernels or as a percentage of the
orthogonal search for larger kernels such as DGEMVT and GESUMMV. The
results show that scalable search produces kernel performance within
1-2\% of the best performance.
%For GEMVER and DGEMV the search space is too large to
%get valuable long running search data.  
%This table shows that for the
%kernels exhaustive data is available the scalable search is within the
%$99^{th}$ percentile in all cases.  The broad coverage search includes
%DGEMVT and GESUMMV and in these cases BTO is within the $98^{th}$
%percentile.

%\begin{table}[h]
%%\caption{Percentile of best version from fast running search as compared to an exhaustive search and a broad coverage search.}
%\caption{Kernel performance for different search strategies.}
%\begin{center}
%\begin{tabular}{|c|c|c|}
%\hline
%\textbf{Kernel} & \textbf{Exhaustive} & \textbf{Broad Coverage} \\ \hline
%\hline
%AXPYDOT & 99 & 100\\ \hline
%VADD & 100 & 100\\ \hline
%WAXPBY & 99 & 100 \\ \hline \hline
%ATAX & 100 & 100 \\ \hline
%BICGK & 100 & 100\\ \hline
%DGEMVT & - & 98 \\ \hline
%GESUMMV & - & 99 \\ \hline
%\end{tabular}
%\end{center}
%\label{tab:percentile}
%\end{table}

\subsection{Evaluation of Search Methods}
\label{sec:compare-strategies}

In the previous sections, we demonstrated that BTO is capable of
generating high-performance routines.  In this section, we examine the
data that led to creating the MFGA search strategy. All of the
experiments in this section were performed on the Intel Westmere.
%% We start with an
%% explanation of the alternative search strategies that we considered.

\subsubsection{Orthogonality of Fusion and Thread Search}
\label{sec:ortho}

The MFGA strategy, for the most part, treats decisions regarding
fusion and thread count orthogonally, which significantly reduces the
size of the search space. However, before we could employ this search
method, we first had to show that it would lead to no degradation in
performance.

We define \emph{orthogonal search} as first searching only the fusion
dimension, then using only the best candidate, searching every viable
thread count.  We evaluated the effectiveness and search time of the
orthogonal search as compared to an exhaustive search using the
smaller kernels: ATAX, AXPYDOT, BICGK, VADD, and WAXPBY.  For all
kernels, orthogonal search found the best-performing version while
taking 1-8\% of the time of exhaustive search, demonstrating that
searching the space orthogonally dramatically reduces search time
without sacrificing performance.  This reduction in search time
results in part from the chosen orthogonal ordering. By searching the
fusion space first, we often dramatically reduce the number of
data-parallel loops and hence the size of the subsequent thread-count
search space.

Thus, we see that fusion and thread search can be conducted
orthogonally without a significant loss of kernel performance.

\subsubsection{Fusion Search}
\label{sec:fuseperf}

Next we focus on fusion strategies.  In this section we analyze our
choice of using a combination of a genetic algorithm and 
the max-fuse heuristic.

We compare four search strategies on our most challenging kernel,
GEMVER. In particular, we test random search, our genetic algorithm
without the max-fuse heuristic, the max-fuse heuristic by itself, and
the combination of the max-fuse heuristic with the genetic algorithm
(MFGA).
As described in Section \ref{sec:search}, the random search strategy
and the genetic algorithm use the same mutation schemes, and thus their
comparison shows the benefit of the crossover and
selection methods.

\begin{figure}[tb] 
  \includegraphics[width=0.5\textwidth]{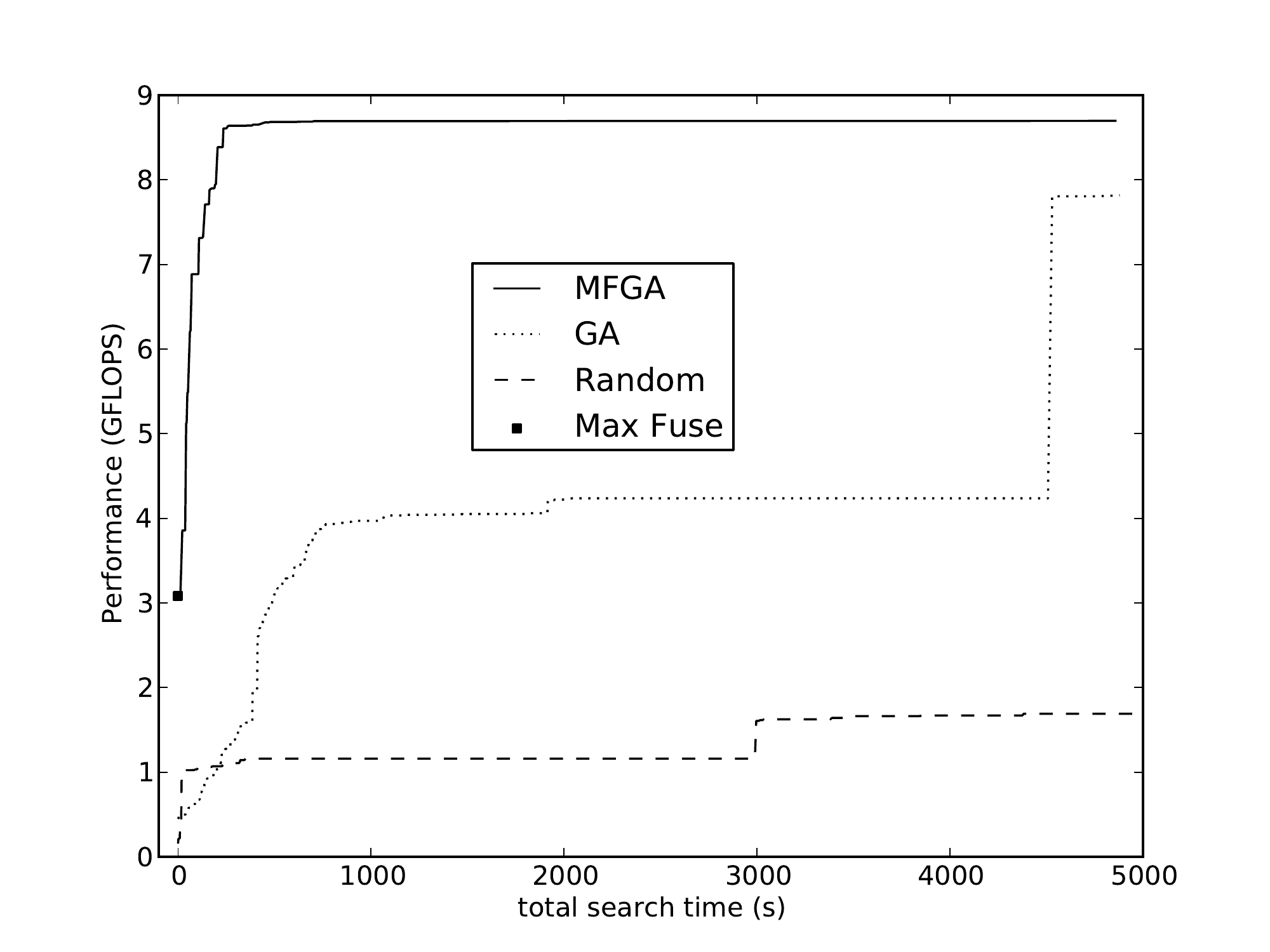} 
  \vspace{-15pt}
   \caption{GEMVER performance over time for different search
     strategies on Intel Westmere. MFGA finds the best version more
     quickly and consistently than either search individually.}
   \label{fig:fusion_search}
\end{figure}

Figure~\ref{fig:fusion_search} shows the performance over time of each
of the search methods. (MF is a single point near 3 GFLOPS.)  Because
the search is stochastic, each of the lines in the chart is the
average of two runs.
MFGA finds the optimal 
point in less than 10 minutes on average.  Without the MF heuristic, 
GA alone eventually reaches 90\% of MFGA but
requires over an hour of search time.  The Random search plateaus without ever
finding the optimal value.  The MF heuristic by itself achieves 40\% of
MFGA.

In conclusion, a combination of GA and MF is the best strategy for 
the fusion portion of the search.

\subsubsection{Thread Search}
\label{sec:threadperf}

Using the MFGA heuristic described in the previous
section, we explore several possible thread search strategies,
including the \emph{global} thread number and the \emph{exhaustive}
strategies discussed in Section~\ref{sec:thread-perf}.
The baseline test is the MFGA search with number of threads set equal
to the number of cores (24 for these experiments), which we refer to
as the \emph{const} strategy.  Recall that the \emph{global} strategy
starts with MFGA and then searches over a single parameter for all
loop nests for the number of threads.  Recall that the
\emph{exhaustive} search replaces the single thread parameter with the
full space of possible thread counts, i.e., considering the number of
threads for each loop nest individually.

\begin{figure}[tb] 
  \includegraphics[width=0.5\textwidth]{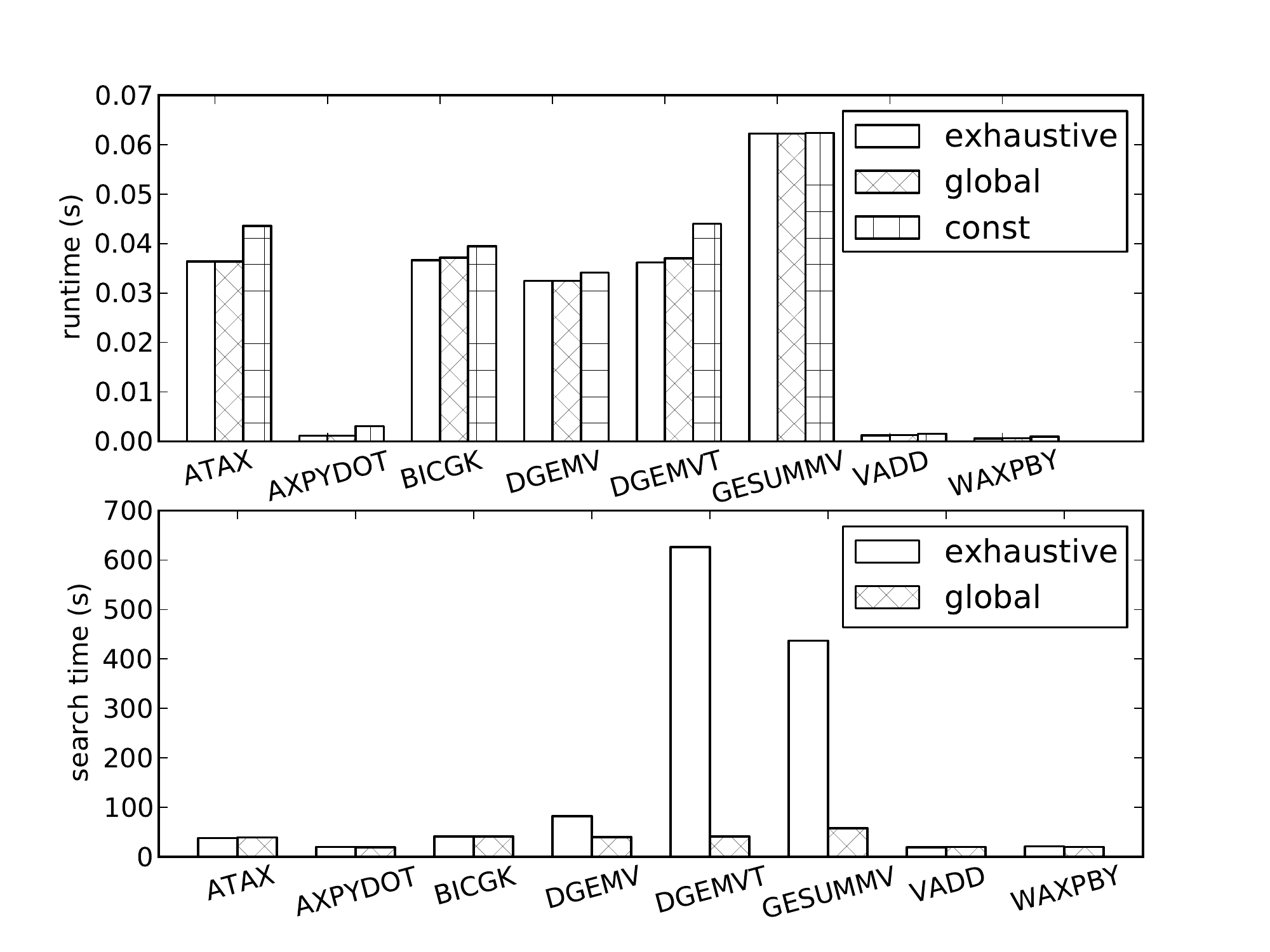} 
   \caption{Best runtime (top) and search time (bottom) for exhaustive and
   global searches. A constant thread number (e.g., equal to the number of cores)
   cannot achieve the runtime
   performance of either global or exhaustive thread search. Searching over a global thread count
   results in a much shorter search time without significantly worsening kernel performance.}
   \label{fig:thread_search}
\end{figure}

%We ran each of the different thread searches for seven different kernels.  
The results for seven kernels are in Figure~\ref{fig:thread_search}.
The top chart shows the final performance of the best version found in
each case.
%As expected, 

Searching over the thread space improves the final performance compared with using 
a constant number of threads (e.g., equal to the number of cores), with negligible
difference in kernel performance between the global thread count (fixed count for all threads) 
and fully exhaustive approaches (varying thread counts for different operations). 
The bottom chart in Figure \ref{fig:thread_search} shows the total search cost of 
the different thread search approaches, demonstrating that global thread 
search improves scalability without sacrificing performance.

%\liz{I can't match the description in the first paragraph with the one in the
%second paragraph or with the legends in the Figure.  The word global is used
%inconsistently, and the word constant only appears in the figure.}

%\begin{figure}[tb] 
  %\includegraphics[width=0.5\textwidth]{threadcost.pdf} 
   %\caption{Total Search Cost for Exhaustive and Global}
   %\label{fig:thread_search1}
%\end{figure}

\section{Related Work} \label{sec:related}

We describe the relationship between our contributions in this paper
and related work in four areas of the literature: loop restructuring
compilers, search strategies for autotuning, partitioning matrix
computations, and empirical search.

% \emph{\textbf{Loop Fusion}}

% The combination of multiple loops into single loop is called loop
% fusion.  When the combined loops access the same data structures, loop
% fusion improves the temporal locality of those accesses.  Applying
% loop fusion to linear algebra computations results in large runtime
% reductions \cite{Howell:2008:CEB:1356052.1356055, Vuduc:2003kl}.
% However, applying loop fusion at every opportunity does not always
% improve performance.  For example, fusing too many loops together can
% result in register spill and extra cache misses \cite{Karlin2011}.

% Example of fusion here?

% Boyana: This is way too broad a subject to cover (if left as compilers), so I'm changing to more specific source-to-source tuning
% I've added more details and moved the treatment of CHiLL and Pluto down below.
% -Jeremy
% CHiLL~\cite{CHiLL} and Pluto~\cite{Pluto} employ a polyhedral model
% to compose loop transformations for optimizing memory hierarchy
% performance and shared-memory parallelization of
% codes.

%\jeremy{The sentence doesn't start with a citation. It starts with
%two names.}
% \emph{\textbf{Domain-specific Compilers}}
% \emph{\textbf{High-Performance Libraries and Autotuning}}

\paragraph*{Loop Fusion and Parallelization}

\citet{Megiddo:1997uq} study the problem of deciding which loops to
fuse in a context where parallelization choices have already been made
(such as an OpenMP program). They model this problem as a weighted
graph whose nodes are loops and whose edges are labeled with the
runtime cost savings resulting from loop fusion.
Because the parallelization choices are
fixed prior to the fusion choices, their approach sometimes misses the
optimal combination of parallelization and fusion decisions.

\citet{Darte:2000vn}, on the other hand, study the space of all fusion
decisions followed by parallelization decisions.
\citet{Pouchet:2010:CIM:1884643.1884672} take a similar approach, they
use a orthogonal approach that exhaustively searches over fusion decisions,
then uses the polyhedral model with analytic models to make tiling and
parallelization decisions.
These approaches roughly correspond to the orthogonal search technique
described in Section~\ref{sec:ortho}.

\citet{Pluto} employs the heuristic of maximally fusing loops. Loop
fusion is generally beneficial, but too much can be detrimental
as it can put too much pressure on registers and
cache~\cite{Karlin2011}.

\citet{Bondhugula:2010kx} develop an analytic model for predicting the
profitability of fusion and parallelization and show speedups relative
to other heuristics such as always fuse and never fuse. However, they
do not validate their model against the entire search space as we do
where possible here.

\paragraph*{Search for Autotuning}

\citet{Vuduc:2004:IJHPCA} study the optimization space of applying
register tiling, loop unrolling, software pipelining, and software
prefetching to matrix multiplication. They show that this search space
is difficult (a very small number of combinations achieve 
good performance), and present a statistical method for
determining when a search has found a point that is close enough to
the best.

\citet{balaprakash2011can} study the effectiveness of several search
algorithms (random search, genetic algorithms, Nelder-Mead simplex) to
find the best combination of optimization decisions from among loop
unrolling, scalar replacement, loop parallelization, vectorization,
and register tilling as implemented in the Orio autotuning 
framework~\cite{Hart2009:Orio}. 
They conclude that the modified Nelder-Mead
method is effective for their search problem. 
%We are currently
%collaborating with these authors to integrate modified Nelder-Mead
%into the BTO compiler.

\citet{CHiLL} develop a framework for empirical search over many loop
optimizations such as permutation, tiling, unroll-and-jam, data
copying, and fusion. They employ an orthogonal search strategy, first
searching over unrolling factors, then tiling sizes, and so on.
\citet{Tiwari2009} describe an autotuning framework that combines
ActiveHarmony's parallel search backend with the CHiLL transformation
framework.

Looptool~\cite{Qasem03improvingperformance} and
AutoLoopTune~\cite{Qasem2006} support loop fusion, unroll-and-jam and
array contraction. AutoLoopTune also supports tiling. POET~\cite{POET}
also supports a number of loop transformations.

\paragraph*{Partitioning Matrix Computations}

The approach to partitioning matrix computations described in this
paper is inspired by the notion of a blocked matrix view in the Matrix
Template Library~\cite{Siek:1999fk}. Several researchers have
subsequently proposed similar abstractions, such as the hierarchically
tiled arrays of \citet{Almasi:2003lr} and the support for matrix
partitioning in FLAME~\cite{Gunnels:2001cr}.

\paragraph*{Search with Empirical Evaluation}

\citet{Bilmes:1997ye} and \citet{Whaley:1998fk} autotune matrix
multiplication using empirical evaluation to determine the
profitability of optimizations.  \citet{Zhao:2005kc} use exhaustive
search and empirical testing to select the best combination of loop
fusion decisions.  \citet{qing08} apply empirical search to determine
the profitability of optimizations for register reuse, SSE
vectorization, strength reduction, loop unrolling, and prefetching.
Their framework is parameterized with respect to the search algorithm
and includes numerous search strategies.

\section{Conclusions and Future Work}
\label{sec:conclusions}

For many problems in high-performance computing, the best solutions require
extensive testing and tuning.  We present an empirical autotuning 
approach for dense matrix algebra that is 
reliable and scalable.  Our tool considers loop fusion, array contraction, 
and shared memory parallelism.

Our experiments have shown that the BTO autotuning system outperforms standard optimizing
compilers and a vendor-optimized BLAS library in most cases, and our results are competitive with
hand-tuned code.  We also describe how we developed our search strategies
and tested the usefulness of each part of the search.

Two big expansions of functionality are planned: distributed memory
support and extension of matrix formats to include triangular, banded, and
sparse.  These extensions will improve the usefulness of BTO, while also
providing an important stress test for the scalability of the search algorithms
and code generation.

% Uncomment this for the final version.

%% \section*{Acknowledgment}

%% %% BN: Argonne ack below, please add yours;
%% This work was supported by the NSF awards CCF 0846121 and CCF
%% 0830458.
%% %
%% This work was also supported by the Office of Advanced
%% Scientific Computing Research, Office of Science, U.S. Dept. of Energy,
%% under Contract DE-AC02-06CH11357. 

%\pagebreak

{%\footnotesize
\bibliographystyle{plainnat}
\bibliography{local}

\begin{thebibliography}{38}
\providecommand{\natexlab}[1]{#1}
\providecommand{\url}[1]{\texttt{#1}}
\expandafter\ifx\csname urlstyle\endcsname\relax
  \providecommand{\doi}[1]{doi: #1}\else
  \providecommand{\doi}{doi: \begingroup \urlstyle{rm}\Url}\fi

\bibitem[Almasi et~al.(2003)Almasi, Rose, Moreira, and Padua]{Almasi:2003lr}
Gheorghe Almasi, Luiz~De Rose, Jose Moreira, and David Padua.
\newblock Programming for locality and parallelism with hierarchically tiled
  arrays.
\newblock In \emph{The 16th International Workshop on Languages and Compilers
  for Parallel Computing}, pages 162--176, College Station, TX, 2003.

\bibitem[Amarasinghe et~al.(2009)Amarasinghe, Campbell, Carlson, Chien, Dally,
  Elnohazy, Hall, Harrison, Harrod, Hill, et~al.]{amarasinghe2009exascale}
S.~Amarasinghe, D.~Campbell, W.~Carlson, A.~Chien, W.~Dally, E.~Elnohazy,
  M.~Hall, R.~Harrison, W.~Harrod, K.~Hill, et~al.
\newblock {Exascale software study: Software challenges in extreme scale
  systems}.
\newblock \emph{DARPA IPTO, Air Force Research Labs, Tech. Rep}, 2009.

\bibitem[Anderson et~al.(1999)Anderson, Gropp, Kaushik, Keyes, and
  Smith]{Anderson}
W.~K. Anderson, W.~D. Gropp, D.~K. Kaushik, D.~E. Keyes, and B.~F. Smith.
\newblock Achieving high sustained performance in an unstructured mesh {CFD}
  application.
\newblock In \emph{Proceedings of the 1999 ACM/IEEE Conference on
  Supercomputing (CDROM)}, Supercomputing '99, Portland, Oregon, 1999. ACM.

\bibitem[Balaprakash et~al.(2011)Balaprakash, Wild, and
  Hovland]{balaprakash2011can}
P.~Balaprakash, S.~Wild, and P.~Hovland.
\newblock Can search algorithms save large-scale automatic performance tuning?
\newblock \emph{Procedia CS}, 4:\penalty0 2136--2145, 2011.

\bibitem[Belter et~al.(2009)Belter, Jessup, Karlin, and Siek]{Belter}
Geoffrey Belter, E.~R. Jessup, Ian Karlin, and Jeremy~G. Siek.
\newblock Automating the generation of composed linear algebra kernels.
\newblock In \emph{SC '09: Proceedings of the Conference on High Performance
  Computing Networking, Storage and Analysis}, pages 1--12, New York, 2009.
  ACM.
\newblock ISBN 978-1-60558-744-8.
\newblock \doi{http://doi.acm.org/10.1145/1654059.1654119}.

\bibitem[Belter et~al.(2010)Belter, Siek, Karlin, and Jessup]{Belter2010}
Geoffrey Belter, Jeremy~G. Siek, Ian Karlin, and E.~R. Jessup.
\newblock Automatic generation of tiled and parallel linear algebra routines.
\newblock In \emph{Fifth International Workshop on Automatic Performance Tuning
  (iWAPT 2010)}, pages 1--15, Berkeley, CA, June 2010.

\bibitem[Bilmes et~al.(1997)Bilmes, Asanovic, Chin, and Demmel]{Bilmes:1997ye}
Jeff Bilmes, Krste Asanovic, Chee-Whye Chin, and Jim Demmel.
\newblock Optimizing matrix multiply using {PHiPAC}: A portable,
  high-performance, {ANSI C} coding methodology.
\newblock In \emph{ICS '97: Proceedings of the 11th International Conference on
  Supercomputing}, pages 340--347, New York, 1997. ACM Press.
\newblock ISBN 0-89791-902-5.
\newblock \doi{http://doi.acm.org/10.1145/263580.263662}.

\bibitem[Blackford et~al.(2002)Blackford, Demmel, Dongarra, Duff, Hammarling,
  Henry, Heroux, Kaufman, Lumsdaine, Petitet, Pozo, Remington, and
  Whaley]{Blackford}
L.~Susan Blackford, James Demmel, Jack Dongarra, Iain Duff, Sven Hammarling,
  Greg Henry, Michael Heroux, Linda Kaufman, Andrew Lumsdaine, Antoine Petitet,
  Roldan Pozo, Karin Remington, and R.~Clint Whaley.
\newblock An updated set of {Basic Linear Algebra Subprograms (BLAS)}.
\newblock \emph{{ACM} Transactions on Mathematical Software}, 28\penalty0
  (2):\penalty0 135--151, June 2002.

\bibitem[Bondhugula et~al.(2008)Bondhugula, Hartono, Ramanujam, and
  Sadayappan]{Pluto}
U.~Bondhugula, A.~Hartono, J.~Ramanujam, and P.~Sadayappan.
\newblock {Pluto}: {A} practical and fully automatic polyhedral program
  optimization system.
\newblock In \emph{Proceedings of the ACM SIGPLAN 2008 Conference on
  Programming Language Design and Implementation (PLDI 08)}, pages 101--113,
  Tucson, AZ, June 2008.

\bibitem[Bondhugula et~al.(2010)Bondhugula, Gunluk, Dash, and
  Renganarayanan]{Bondhugula:2010kx}
Uday Bondhugula, Oktay Gunluk, Sanjeeb Dash, and Lakshminarayanan
  Renganarayanan.
\newblock A model for fusion and code motion in an automatic parallelizing
  compiler.
\newblock In \emph{Proceedings of the 19th International Conference on Parallel
  Architectures and Compilation Techniques}, PACT~'10, pages 343--352, New
  York, 2010. ACM.

\bibitem[Chen et~al.(2008)Chen, Chame, and Hall]{CHiLL}
C.~Chen, J.~Chame, and M.~Hall.
\newblock {CHiLL}: {A} framework for composing high-level loop transformations.
\newblock Technical Report 08-897, Department of Computer Science, University
  of Southern California, 2008.

\bibitem[Dagum and Menon(1998)]{Dagum:1998:OIA:615255.615542}
Leonardo Dagum and Ramesh Menon.
\newblock Openmp: An industry-standard {API} for shared-memory programming.
\newblock \emph{IEEE Comput. Sci. Eng.}, 5\penalty0 (1):\penalty0 46--55,
  January 1998.
\newblock ISSN 1070-9924.
\newblock \doi{10.1109/99.660313}.
\newblock URL \url{http://dx.doi.org/10.1109/99.660313}.

\bibitem[Darte and Huard(2000)]{Darte:2000vn}
Alain Darte and Guillaume Huard.
\newblock Loop shifting for loop parallelization.
\newblock Technical Report 2000-22, Ecole Normale Superieure de Lyon, May 2000.

\bibitem[Dongarra et~al.(1988)Dongarra, Croz, Hammarling, and
  Hanson]{Dongarra88}
Jack~J. Dongarra, Jeremy~De Croz, Sven Hammarling, and Richard~J. Hanson.
\newblock An extended set of {FORTRAN Basic Linear Algebra Subprograms}.
\newblock \emph{{ACM} Transactions on Mathematical Software}, 14\penalty0
  (1):\penalty0 1--17, March 1988.

\bibitem[Dongarra et~al.(1990)Dongarra, Croz, Hammarling, and Duff]{Dongarra90}
Jack~J. Dongarra, Jeremy~Du Croz, Sven Hammarling, and Iain Duff.
\newblock A set of level 3 {Basic Linear Algebra Subprograms}.
\newblock \emph{{ACM} Transactions on Mathematical Software}, 16\penalty0
  (1):\penalty0 1--17, March 1990.

\bibitem[Gunnels et~al.(2001)Gunnels, Gustavson, Henry, and van~de
  Geijn]{Gunnels:2001cr}
John~A. Gunnels, Fred~G. Gustavson, Greg~M. Henry, and Robert~A. van~de Geijn.
\newblock {FLAME}: Formal linear algebra methods environment.
\newblock \emph{ACM Trans. Math. Softw.}, 27\penalty0 (4):\penalty0 422--455,
  2001.

\bibitem[Hartono et~al.(2009)Hartono, Norris, and Sadayappan]{Hart2009:Orio}
Albert Hartono, Boyana Norris, and Ponnuswamy Sadayappan.
\newblock Annotation-based empirical performance tuning using {Orio}.
\newblock In \emph{IPDPS '09: Proceedings of the 2009 IEEE International
  Symposium on Parallel \& Distributed Processing}, pages 1--11, Washington,
  DC, 2009. IEEE Computer Society.
\newblock ISBN 978-1-4244-3751-1.
\newblock \doi{http://dx.doi.org/10.1109/IPDPS.2009.5161004}.
\newblock URL \url{http://www.mcs.anl.gov/uploads/cels/papers/P1556.pdf}.
\newblock {A}lso available as Preprint ANL/MCS-P1556-1008.

\bibitem[Howell et~al.(2008)Howell, Demmel, Fulton, Hammarling, and
  Marmol]{Howell:2008:CEB:1356052.1356055}
Gary~W. Howell, James~W. Demmel, Charles~T. Fulton, Sven Hammarling, and Karen
  Marmol.
\newblock Cache efficient bidiagonalization using {BLAS} 2.5 operators.
\newblock \emph{ACM Trans. Math. Softw.}, 34:\penalty0 14:1--14:33, May 2008.

\bibitem[Intel(2012)]{icc}
Intel.
\newblock Intel {C}omposer.
\newblock \url{http://software.intel.com/en-us/articles/intel-compilers}, April
  2012.

\bibitem[Karlin et~al.(2011{\natexlab{a}})Karlin, Jessup, Belter, and
  Siek]{Karlin:2011:PMP:1964218.1964226}
Ian Karlin, Elizabeth Jessup, Geoffrey Belter, and Jeremy~G. Siek.
\newblock Parallel memory prediction for fused linear algebra kernels.
\newblock \emph{SIGMETRICS Perform. Eval. Rev.}, 38:\penalty0 43--49, March
  2011{\natexlab{a}}.
\newblock ISSN 0163-5999.
\newblock \doi{http://doi.acm.org/10.1145/1964218.1964226}.
\newblock URL \url{http://doi.acm.org/10.1145/1964218.1964226}.

\bibitem[Karlin et~al.(2011{\natexlab{b}})Karlin, Jessup, and
  Silkensen]{Karlin2011}
Ian Karlin, Elizabeth Jessup, and Erik Silkensen.
\newblock Modeling the memory and performance impacts of loop fusion.
\newblock \emph{Journal of Computational Science}, In press,
  2011{\natexlab{b}}.
\newblock ISSN 1877-7503.
\newblock \doi{DOI: 10.1016/j.jocs.2011.03.002}.

\bibitem[Karp et~al.(1967)Karp, Miller, and
  Winograd]{Karp:1967:OCU:321406.321418}
Richard~M. Karp, Raymond~E. Miller, and Shmuel Winograd.
\newblock The organization of computations for uniform recurrence equations.
\newblock \emph{J. ACM}, 14\penalty0 (3):\penalty0 563--590, July 1967.
\newblock ISSN 0004-5411.
\newblock \doi{10.1145/321406.321418}.
\newblock URL \url{http://doi.acm.org/10.1145/321406.321418}.

\bibitem[Lawson et~al.(1979)Lawson, Hanson, Kincaid, and Krogh]{Lawson}
C.~L. Lawson, R.~J. Hanson, D.~R. Kincaid, and F.~T. Krogh.
\newblock {Basic Linear Algebra Subprograms} for {Fortran} usage.
\newblock \emph{{ACM} Transactions on Mathematical Software}, 5\penalty0
  (3):\penalty0 308--323, September 1979.

\bibitem[Megiddo and Sarkar(1997)]{Megiddo:1997uq}
Nimrod Megiddo and Vivek Sarkar.
\newblock Optimal weighted loop fusion for parallel programs.
\newblock In \emph{Proceedings of the Ninth Annual ACM Symposium on Parallel
  Algorithms and Architectures}, SPAA '97, pages 282--291, New York, 1997. ACM.

\bibitem[Mitchell(1998)]{mitchell1998introduction}
M.~Mitchell.
\newblock \emph{An introduction to genetic algorithms}.
\newblock The MIT Press, 1998.
\newblock ISBN 0262631857.

\bibitem[Mueller(1999)]{Mueller94pthreadslibrary}
Frank Mueller.
\newblock Pthreads library interface.
\newblock Technical report, Florida State University, 1999.

\bibitem[{Portland Group}(2012)]{pgi}
{Portland Group}.
\newblock Portland group compiler.
\newblock \url{http://www.pgroup.com}, April 2012.

\bibitem[Pouchet et~al.(2010)Pouchet, Bondhugula, Bastoul, Cohen, Ramanujam,
  and Sadayappan]{Pouchet:2010:CIM:1884643.1884672}
Louis-No\"{e}l Pouchet, Uday Bondhugula, C\'{e}dric Bastoul, Albert Cohen,
  J.~Ramanujam, and P.~Sadayappan.
\newblock Combined iterative and model-driven optimization in an automatic
  parallelization framework.
\newblock In \emph{Proceedings of the 2010 ACM/IEEE International Conference
  for High Performance Computing, Networking, Storage and Analysis}, SC '10,
  pages 1--11, Washington, DC, November 2010. IEEE Computer Society.

\bibitem[Qasem et~al.(2003)Qasem, Jin, and
  Mellor-{Crummey}]{Qasem03improvingperformance}
Apan Qasem, Guohua Jin, and John Mellor-{Crummey}.
\newblock Improving performance with integrated program transformations.
\newblock Technical Report TR03-419, Department of Computer Science, Rice
  University, October 2003.

\bibitem[Qasem et~al.(2006)Qasem, Kennedy, and Mellor-{Crummey}]{Qasem2006}
Apan Qasem, Ken Kennedy, and John Mellor-{Crummey}.
\newblock Automatic tuning of whole applications using direct search and a
  performance-based transformation system.
\newblock \emph{The Journal of Supercomputing: Special Issue on Computer
  Science Research Supporting High-Performance Applications}, 36\penalty0
  (9):\penalty0 183--196, May 2006.

\bibitem[Siek(1999)]{Siek:1999fk}
Jeremy~G. Siek.
\newblock A modern framework for portable high performance numerical linear
  algebra.
\newblock Master's thesis, University of Notre Dame, 1999.

\bibitem[Siek et~al.(2008)Siek, Karlin, and Jessup]{Siek}
Jeremy~G. Siek, Ian Karlin, and E.~R. Jessup.
\newblock Build to order linear algebra kernels.
\newblock In \emph{Workshop on {P}erformance {O}ptimization for {H}igh-{L}evel
  {L}anguages and {L}ibraries ({POHLL} 2008)}, pages 1--8, Miami, FL, April
  2008.

\bibitem[Tiwari et~al.(2009)Tiwari, Chen, Chame, Hall, and
  Hollingsworth]{Tiwari2009}
Ananta Tiwari, Chun Chen, Jacqueline Chame, Mary Hall, and Jeffrey~K.
  Hollingsworth.
\newblock A scalable autotuning framework for compiler optimization.
\newblock In \emph{Proceedings of the 23rd IEEE International Parallel \&
  Distributed Processing Symposium}, Rome, Italy, May 2009.

\bibitem[Vuduc et~al.(2004)Vuduc, Demmel, and Bilmes]{Vuduc:2004:IJHPCA}
Richard Vuduc, James~W. Demmel, and Jeff~A. Bilmes.
\newblock Statistical models for empirical search-based performance tuning.
\newblock \emph{International Journal of High Performance Computing
  Applications}, 18\penalty0 (1):\penalty0 65--94, 2004.
\newblock \doi{10.1177/1094342004041293}.
\newblock URL \url{http://hpc.sagepub.com/content/18/1/65.abstract}.

\bibitem[Whaley and Dongarra(1998)]{Whaley:1998fk}
R.~Clint Whaley and Jack~J. Dongarra.
\newblock Automatically tuned linear algebra software.
\newblock In \emph{Supercomputing '98: Proceedings of the 1998 ACM/IEEE
  conference on Supercomputing (CDROM)}, pages 1--27, Washington, DC, 1998.
  IEEE Computer Society.
\newblock ISBN 0-89791-984-X.

\bibitem[Yi and Qasem(2008)]{qing08}
Qing Yi and Apan Qasem.
\newblock Exploring the optimization space of dense linear algebra kernels.
\newblock In \emph{Languages and Compilers for Parallel Computing: 21th
  International Workshop, LCPC 2008, Edmonton, Canada, July 31 - August 2,
  2008, Revised Selected Papers}, pages 343--355, Berlin, 2008.
  Springer-Verlag.
\newblock ISBN 978-3-540-89739-2.
\newblock \doi{http://dx.doi.org/10.1007/978-3-540-89740-8_24}.

\bibitem[Yi et~al.(2007)Yi, Seymour, You, Vuduc, and Quinlan]{POET}
Qing Yi, Keith Seymour, Haihang You, Richard Vuduc, and Dan Quinlan.
\newblock {POET}: {P}arameterized optimizations for empirical tuning.
\newblock In \emph{Proceedings of the Parallel and Distributed Processing
  Symposium, 2007}, pages 1--8, Long Beach, CA, March 2007. IEEE.
\newblock \doi{10.1109/IPDPS.2007.370637}.

\bibitem[Zhao et~al.(2005)Zhao, Yi, Kennedy, Quinlan, and Vuduc]{Zhao:2005kc}
Y.~Zhao, Q.~Yi, K.~Kennedy, D.~Quinlan, and R.~Vuduc.
\newblock Parameterizing loop fusion for automated empirical tuning.
\newblock Technical Report UCRL-TR-217808, Center for Applied Scientific
  Computing, Lawrence Livermore National Laboratory, December 2005.

\end{thebibliography}
}

\end{document}